\documentclass[11pt,a4paper]{article}
\usepackage{comment}
\usepackage{graphicx}
\usepackage{amssymb}
\usepackage{amsmath}
\usepackage{amsfonts}
\usepackage{dsfont}
\usepackage{mathtools}
\usepackage{array}
\usepackage{soul}
\usepackage{float}
\usepackage{rotating}
\usepackage{bbold,amsfonts}
\allowdisplaybreaks

\usepackage[utf8]{inputenc}
\usepackage{bm}
\usepackage{xcolor}
\usepackage{float}
\usepackage{braket}
\usepackage{placeins}
\usepackage[height=8.8in,width=6.45in]{geometry}
\usepackage[font=small,labelfont=bf]{caption}
\usepackage[hidelinks]{hyperref}
\usepackage{booktabs,float,slashed}
\usepackage{standalone}
\usepackage{caption}
\usepackage{subcaption}
\usepackage{makecell}
\usepackage[maxbibnames=99,sorting=none,giveninits=true,backend=biber,style=numeric-comp,sortcites,doi=false,hyperref=true]{biblatex}
\renewbibmacro{in:}{}
\usepackage{hyperref}
\DeclareFieldFormat{doilink}{\iffieldundef{doi}{#1}{\href{https://doi.org/\thefield{doi}}{#1}}}
\DeclareFieldFormat[article,periodical]{volume}{\mkbibbold{#1}}
\DeclareFieldFormat[article,periodical]{journaltitle}{#1}
\DeclareFieldFormat[article,periodical]{pages}{#1}

\usepackage{xpatch}
\xpatchbibdriver{article}
  {\usebibmacro{journal+issuetitle}%
   \newunit
   \usebibmacro{byeditor+others}%
   \newunit
   \usebibmacro{note+pages}}
  {\printtext[doilink]{%
     \usebibmacro{journal+issuetitle}%
     \newunit
     \usebibmacro{byeditor+others}%
     \newunit
     \usebibmacro{note+pages}}}
  {}{}

\numberwithin{equation}{section}

\newcommand{\beq}{\begin{equation}}
\newcommand{\eeq}{\end{equation}}

\makeatletter
\newcommand*{\letterdef@}{}
\newcommand*{\letterdef}[3]{%
	\def\letterdef@##1{\expandafter\newcommand\csname #1\endcsname{#2{##1}}}%
	\@tfor\@tempa :=#3\do{\expandafter\letterdef@\expandafter{\@tempa}}}
\makeatother
\letterdef{c#1} {\mathcal}{ABCDEFGHIJKLMNOPQRSTUVWXYZ} 
\letterdef{rm#1}{\mathrm} {dDeimM} 

\title{Draft-Wilson lines in $Z_2$ quiver theory}

\begin{document}

\begin{titlepage}

\begin{flushright}
\small
\texttt{HU-EP-25/23}
\end{flushright}

\vspace*{10mm}
\begin{center}
{\LARGE \bf Integrated correlators of coincident Wilson lines\\
in $SU(N)$ gauge theories at strong coupling
}

\vspace*{15mm}

{\Large Lorenzo De Lillo${}^{\,a,b}$ and  Alessandro Pini${}^{\,c}$}

\vspace*{8mm}

${}^a$ Universit\`a di Torino, Dipartimento di Fisica,\\
			Via P. Giuria 1, I-10125 Torino, Italy
			\vskip 0.3cm
			
${}^b$   I.N.F.N. - sezione di Torino,\\
			Via P. Giuria 1, I-10125 Torino, Italy 
			\vskip 0.3cm

${}^c$ Institut f{\"u}r Physik, Humboldt-Universit{\"a}t zu Berlin,\\
     IRIS Geb{\"a}ude, Zum Großen Windkanal 2, 12489 Berlin, Germany  
     \vskip 0.3cm

\vskip 0.8cm
	{\small
		E-mail:
		\texttt{lorenzo.delillo@unito.it;alessandro.pini@physik.hu-berlin.de}
	}
\vspace*{0.8cm}
\end{center}

\begin{abstract}
We consider two four-dimensional gauge theories with gauge group $SU(N)$: the $\mathcal{N}=4$ Super Yang-Mills (SYM) theory  and the $\mathcal{N}=2$ quiver gauge theory obtained as a $\mathbb{Z}_2$ orbifold of $\mathcal{N}=4$ SYM. In this context, we study a novel class of integrated correlators, namely those involving $n$-coincident Wilson lines and two moment map operators of conformal dimension two. By exploiting supersymmetric localization, we obtain exact expressions for these observables valid in the large-$N$ limit. Furthermore, using a combination of analytical and numerical methods, we derive their strong coupling expansions.
\end{abstract}

\end{titlepage}
\setcounter{tocdepth}{2}

\newpage

\tableofcontents

\vspace*{1cm}

\section{Introduction}
In recent years, the study of integrated correlators has attracted growing interest. These observables have proven to be powerful tools for probing non-perturbative aspects of supersymmetric gauge theories and their strong coupling regime. In particular, they have been extensively investigated in the context of the $\mathcal{N}=4$ Super Yang-Mills (SYM) theory, where the primary focus has been on integrated correlators of four local scalar operators $\mathcal{O}_p$ belonging to the stress-tensor multiplet and transforming in the $[0,p,0]$ representation of the $SU(4)$ $R$-symmetry group, with two of them taken to have conformal dimension $p=2$ \cite{Binder:2019jwn,Chester:2019jas,Chester:2020dja,Chester:2020vyz,Dorigoni:2021bvj,Dorigoni:2021guq}. These observables are constructed by integrating the four-point function among the $\mathcal{O}_p$ operators, which depends on two conformal cross-ratios, over the spacetime coordinates with specific integration measures determined uniquely by superconformal invariance. Remarkably, they can be computed exactly by exploiting supersymmetric localization \cite{Pestun:2007rz}. This is achieved by placing the $\mathcal{N}=4$ SYM theory on a unit four-sphere $\mathbb{S}^4$ and deforming it to the $\mathcal{N}=2^{*}$ theory, where the adjoint hypermultiplet acquires a mass $m$. The integrated correlators are then obtained by taking suitable derivatives of the mass-deformed partition function with respect to the mass and to the couplings $\tau_p$ and $\bar{\tau}_p$ associated with the operators $\mathcal{O}_p$ and $\overline{\mathcal{O}}_p$, respectively, and finally setting the mass to zero.

The study of these observables has been further developed in recent works \cite{Dorigoni:2022zcr,Wen:2022oky,Dorigoni:2023ezg,Brown:2023zbr,Brown:2024tru,Zhang:2024ypu,Kim:2024pjb}, and several of their properties have been elucidated through the analysis of the insertion of operators with large conformal dimensions \cite{Brown:2023cpz,Brown:2023why,Paul:2023rka} and the investigation of their modular properties \cite{Chester:2019pvm,Alday:2021vfb,Dorigoni:2022cua,Alday:2023pet,Dorigoni:2024dhy}. Moreover, the extension of these observables to $\mathcal{N}=2$ supersymmetric theories has also been investigated \cite{Chester:2022sqb,Fiol:2023cml,Behan:2023fqq,Billo:2023kak,Pini:2024uia,Billo:2024ftq,Chester:2025ssu}.

More recently, a different type of integrated correlator, involving a defect operator (such as a 1/2 BPS Wilson line) together with two scalar operators $\mathcal{O}_2$, has been considered. Also this observable can be computed exactly using supersymmetric localization, by taking two mass derivatives of the vacuum expectation value of the Wilson loop in the mass-deformed $\mathcal{N}=2^{*}$ theory and then setting the mass to zero \cite{Pufu:2023vwo}. In this case as well, the corresponding unintegrated correlation function depends on two conformal cross-ratios and the coupling constant \cite{Buchbinder:2012vr}. The explicit form of the integration measure for this correlator was derived in \cite{Billo:2023ncz,Billo:2024kri,Dempsey:2024vkf} by analyzing the residual superconformal symmetry in the presence of the defect and exploiting the set of superconformal Ward identities of the $\mathcal{N}=2^{*}$ theory. The study of this latter type of integrated correlator is particularly relevant, as line operators generally transform non-trivially under the $\mathcal{S}$-duality group of $\mathcal{N}=4$ SYM \cite{MONTONEN1977117}. Exploiting this property, the modular behavior of these correlators has been analyzed in \cite{Dorigoni:2024vrb,Dorigoni:2024csx}. 
More recently, this observable has also been studied in $\mathcal{N}=4$ SYM with $\mathrm{Sp}(N)$ gauge group~\cite{DeLillo:2025hal}, as well as in certain $\mathcal{N}=2$ gauge theories~\cite{DeLillo:2025hal,Pini:2024zwi,DeSmet:2025mbc}, where the local operators entering the correlator are now given by the so-called “moment map operators”, i.e., scalar chiral operators of conformal dimension two belonging to the $\hat{\mathcal{B}}_1$ multiplet of the $\mathfrak{su}(2,2|2)$ superconformal algebra.

In this work, we initiate the study of a generalization of the previously discussed integrated correlator, where the defect operator consists of $n$ coincident Wilson lines. It is important to note that the residual superconformal symmetry preserved by this new type of defect coincides with that of the $n=1$ case. Moreover, since this defect operator remains 1/2 BPS, an explicit expression for the integration measure can still be derived by exploiting the same set of Ward identities employed in \cite{Billo:2023ncz,Billo:2024kri,Dempsey:2024vkf}. Crucially for the present work, this observable can still be computed exactly via supersymmetric localization. For instance, in the case of $\mathcal{N}=4$ SYM theory, it is given by
\begin{align}
\mathcal{I}^{(n)}_{\mathcal{N}=4} = \partial^2_{m} \langle (W)^n \rangle \Big|_{m=0} \,,
\label{IntegratedCorrelatorN4}
\end{align}
where $\langle (W)^n \rangle$ denotes the vacuum expectation value of the $n$ coincident Wilson lines in the mass-deformed $\mathcal{N}=2^{*}$ theory. The aim of this paper is twofold. First, we focus on the $\mathcal{N}=4$ SYM theory with gauge group $SU(N)$ and initiate a systematic study of the correlator \eqref{IntegratedCorrelatorN4} in the large-$N$ limit. Additionally, we extend our analysis to $\mathcal{N}=2$ superconformal field theories and, to this end, we consider the two-node quiver gauge theory obtained as a $\mathbb{Z}_2$ orbifold projection of $\mathcal{N}=4$ SYM.  We restrict our attention to the orbifold fixed point of the theory, where the two Yang-Mills couplings are set equal. This choice is motivated by the fact that, at this symmetric point, the theory admits a known gravity dual given by $AdS_5 \times S^5/\mathbb{Z}_2$ \cite{Kachru:1998ys,Gukov:1998kk}. Furthermore, one of the major technical challenges typically encountered in non-maximally supersymmetric theories, namely the non-Gaussian nature of the matrix model arising from localization, in this case can be effectively addressed. Indeed, recent developments \cite{Beccaria:2021hvt,Billo:2021rdb,Billo:2022fnb} have introduced techniques enabling the exact computation of various observables in the large-$N$ limit of this theory. These include, among others, 2- and 3-point functions of chiral primary operators \cite{Galvagno:2020cgq,Billo:2022xas,Billo:2022gmq,Beccaria:2022ypy,Billo:2022lrv,Korchemsky:2025eyc,} and expectation values of Wilson loop correlators \cite{Beccaria:2021ksw,,Galvagno:2021bbj,Preti:2022inu,Beccaria:2023kbl,Pini:2023lyo}. It is therefore natural to investigate whether these techniques can also be successfully applied to the computation of the integrated correlator considered in this work.

It is important to note that the aforementioned $\mathcal{N}=2$ quiver gauge theory differs from $\mathcal{N}=4$ SYM in one significant physical aspect. Specifically, the quiver theory is invariant under a discrete $\mathbb{Z}_2$ symmetry that exchanges the two gauge nodes. As a consequence, one can construct observables that are either even or odd under this symmetry. This distinction, in particular, applies to Wilson line operators, for which one can define twisted ($+$) and untwisted ($-$) linear combinations \cite{Rey:2010ry,Galvagno:2021bbj}
\begin{align}
W{\pm} = \frac{1}{\sqrt{2}}(W_0 \pm W_1) \, ,
\end{align}
where $W_0$ and $W_1$ denote Wilson loops in the fundamental representation of the first and second gauge nodes, respectively. Within the context of the $\mathcal{N}=2$ quiver theory, we can thus consider integrated correlators involving $n_1$ untwisted and $2n_2$ twisted Wilson loops\footnote{Correlators involving an odd number of twisted Wilson loops trivially vanish due to the $\mathbb{Z}_2$ symmetry.} and, using supersymmetric localization, these observables can be computed as
\begin{align}
\mathcal{I}^{(n_1,n_2)} = \partial^2_{m_i}\,\langle \left(W_{+}\right)^{n_1}\, \left(W_{-}\right)^{2n_2} \rangle \Big|_{m_i=0} \, ,
\label{IntegratedCorrelatorN2}
\end{align}
where the vacuum expectation value is evaluated in the mass-deformed theory in which the two matter hypermultiplets are given distinct masses $m_1$ and $m_2$. As a final remark, we observe that correlators involving only twisted Wilson loops do not have a counterpart in $\mathcal{N}=4$ SYM and therefore constitute genuinely new observables, characteristic of the $\mathcal{N}=2$ quiver theory.

The rest of this article is organized as follows. In Section \ref{sec:MatrixModel}, we review the main properties of the two theories considered in this work, introduce the corresponding mass-deformed matrix models and explain how computations can be efficiently performed in the large-$N$ limit. Then, in Section \ref{sec:N4}, we focus on the maximal supersymmetric theory and, after a long computation, derive the first three orders of the large-$N$ expansion of the integrated correlator~\eqref{IntegratedCorrelatorN4}. We then analytically evaluate these expressions in the strong coupling regime. In Section \ref{sec:IntegratedCorrelatorZ2}, we turn to the integrated correlator \eqref{IntegratedCorrelatorN2} in the $\mathbb{Z}_2$ quiver theory, computing its large-$N$ expansion  both at finite 't Hooft coupling and in the large-$\lambda$ limit, and discuss similarities and differences with respect to the $\mathcal{N}=4$ SYM case. Our conclusions are presented in Section \ref{sec:conclusions}.  More technical details regarding the analytical and numerical methods used to derive the strong coupling expansions, as well as some properties of $\mathcal{N}=4$ SYM at large $N$ needed for obtaining our results, are collected in three Appendices. Finally, some particularly lengthy $\mathcal{N}=4$ expressions are collected in an auxiliary \texttt{Mathematica} file.

\section{The gauge theories and their mass deformations}
\label{sec:MatrixModel}
\begin{figure}[h!]
  \centering
\includegraphics[width=0.45\textwidth]{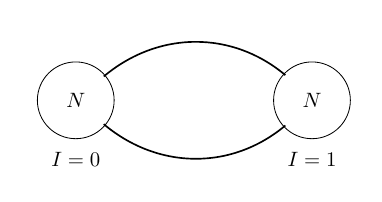}
  \caption{The $4d$ $\mathcal{N}=2$ two nodes quiver gauge theory.}
  \label{fig:QuiverZ2}
\end{figure}

In this work we consider the $\mathcal{N}=4$ SYM theory with gauge group $SU(N)$ and the $\mathcal{N}=2$ quiver gauge theory arising as a $\mathbb{Z}_2$ orbifold of $\mathcal{N}=4$ SYM with gauge group $SU(2N)$. It is convenient to summarize the matter content of the latter with the quiver diagram reported in Figure \ref{fig:QuiverZ2},
where each node represents an $SU(N)$ gauge group and each line corresponds to an $\mathcal{N}=2$ hypermultiplet in the bifundamental representation. Given this matter content, one can actually show that the $\beta$-function at each node vanishes, thereby making the theory conformal. Moreover, we only concentrate on the so-called ``orbifold fixed point", where the coupling constants associated with the two nodes labeled by the index $I=0,1$ are set equal to a common value, namely $g_{0}=g_{1}=g$.

\subsection{The matrix models}
In order to evaluate the integrated correlators \eqref{IntegratedCorrelatorN4} and \eqref{IntegratedCorrelatorN2} we have to consider a proper mass deformation of the aforementioned theories. In particular, in the case of $\mathcal{N}=4$ SYM, we give a mass $m$ to the adjoint hypermultiplet, obtaining the so-called $\mathcal{N}=2^*$ theory, while, in the case of the $\mathcal{N}=2$ quiver gauge theory, we assign masses $m_1$ and $m_2$ to the two bifundamental hypermultiplets. By subsequently placing these theories on a unit 4-sphere $\mathbb{S}^4$, we can exploit supersymmetric localization \cite{Pestun:2007rz} to rewrite the corresponding mass-dependent partition functions $\mathcal{Z}_{\mathcal{N}=2^*}$ and $\mathcal{Z}(m_1,m_2)$ as 
\begin{align}
&  \mathcal{Z}_{\mathcal{N}=2^*}(m)= \left(\frac{8\pi^2N}{\lambda} \right)^{N^2-1}\int da \; \text{e}^{-\frac{8\pi^2N}{\lambda}\text{tr}\,a^2}\, \Big|\mathcal{Z}_{\text{1-loop}}(a,m)\,\mathcal{Z}_{\text{inst}}(a,m)\Big|^2\, \;,
\label{ZN2star}
\\[0.5em]
& \mathcal{Z}(m_1,m_2)= \left(\frac{8\pi^2N}{\lambda} \right)^{N^2-1}\int da \int db \; \text{e}^{-\frac{8\pi^2N}{\lambda}(\text{tr}\,a^2 +\text{tr}\,b^2 )}\,  \Big|\mathcal{Z}_{\text{1-loop}}(a,b,m_1,m_2)\,\mathcal{Z}_{\text{inst}}(a,b,m_1,m_2)\Big|^2\, \ ,
\label{Zquiver}
\end{align}
where $\lambda = g^2\,N$ is the 't Hooft coupling, while $a,b$ denote matrices belonging to the $\mathfrak{su}(N)$ Lie algebra \footnote{These two matrices and the integration measures $da$ and $db$, can be expressed in terms of the generators $ T_c$ of the $\mathfrak{su}(N)$ Lie algebra, which we take to be normalized as $\text{tr}(T_b T_c) = \frac{1}{2} \delta_{b,c}$, in the following way
\begin{align}
a = \sum_{c=1}^{N^2-1}a_c\,T^c\,, \quad b = \sum_{c=1}^{N^2-1}b_c\,T^c\, , \qquad da = \prod_{c=1}^{N^2-1}\frac{da_c}{\sqrt{2\pi}}\, , \quad db = \prod_{c=1}^{N^2-1}\frac{db_c}{\sqrt{2\pi}} \,\ .    
\end{align}
}. $\mathcal{Z}_{\text{1-loop}}$ and $\mathcal{Z}_{\text{inst}}$ represent the 1-loop and instanton contributions respectively. In the large-$N$ limit we can neglect this last term since it is exponentially suppressed, therefore we simply set it to 1. Instead, $\mathcal{Z}_{1-\text{loop}}$ depends on the matter content of the theory and, for our purposes, we need to consider its small-mass expansion up to the quadratic order. This analysis was already performed in \cite{Billo:2023ncz} and \cite{Pini:2024uia} for the $\mathcal{N}=2^*$ theory and the mass-deformed quiver gauge theory, respectively. Therefore, here we simply report the corresponding results. In particular, after performing the rescalings
\begin{align}
a \mapsto \sqrt{\frac{\lambda}{8\pi^2N}}a, \qquad b \mapsto \sqrt{\frac{\lambda}{8\pi^2N}}b\, \ , 
\end{align}
for the $\mathcal{N}=2^*$ theory one obtains \cite{Billo:2023ncz}  
\begin{align}
\mathcal{Z}_{\mathcal{N}=2^{\star}}(m) = \int da \,\text{exp}\left[\displaystyle -\text{tr}a^2+m^2\,\mathcal{M}_{0}+O(m^4)\right]\, \ ,
\end{align}
where 
\begin{align}
&\mathcal{M}_{0} =   - \sum_{n=1}^{\infty}\sum_{\ell=0}^{2n}(-1)^{\ell+n}\frac{\zeta_{2n+1}(2n+1)!}{\ell!(2n-\ell)!} \bigg( \frac{\lambda}{8 \pi^2 N}
 \bigg)^{n}\text{tr} a^{2n-\ell}\text{tr} a^\ell  \;, \label{M0}
\end{align}
with $\zeta_{n}$ denotes the value of the Riemann zeta function $\zeta(n)$. Instead, the small mass expansion of the partition function \eqref{Zquiver} reads \cite{Pini:2024uia}
\begin{align}
\mathcal{Z}(m_1,m_2) = \int da \int db\,\text{exp}\left[\displaystyle -(\text{tr}a^2+ \text{tr}b^2)-S_0+(m_1^2+m_2^2)\,\mathcal{M}_{\mathbb{Z}_2}+O(m^4)\right]\, \ ,
\end{align}
where \cite{Galvagno:2021bbj}
\begin{align}
&S_0 = \sum_{m=2}^{\infty}\sum_{k=2}^{2m}(-1)^{m+k} \left(\frac{\lambda}{8\pi^2N}\right)^m \binom{2m}{k} \frac{\zeta_{2m-1}}{m}\left(\text{tr}a^{2m-k}-\text{tr}b^{2m-k} \right) \left(\text{tr}a^k-\text{tr}b^k \right) \, ,\label{S0}\\
& \mathcal{M}_{\mathbb{Z}_2}= - \sum_{n=1}^{\infty}\sum_{\ell=0}^{2n}(-1)^{n+\ell}\frac{(2n+1)!\zeta_{2n+1}}{(2n-\ell)! \ell!} \left(\frac{\lambda}{8\pi^2N} \right)^n \;\text{tr}a^\ell  \;\text{tr}b^{2n-\ell} \;. \label{MZ2}
\end{align}

We are now ready to introduce the matrix model representation of the half-BPS circular Wilson loop in the fundamental representation. In the case of  $\mathcal{N}=4$ SYM theory it is given by
\cite{Pestun:2007rz, Beccaria:2021ism}
\begin{align}
W_0(a,\lambda) =   \frac{1}{N}\sum_{k=0}^{\infty}\frac{1}{k!}\left(\frac{\lambda}{2N}\right)^{\frac{k}{2}} \text{tr}a^{k} \;.
\label{WN4}
\end{align}
Consequently, the small mass expansion of the vacuum expectation value of $n$ coincident Wilson loops in the $\mathcal{N}=2^{*}$ theory takes the following form
\begin{align}
& \mathcal{W}^{(n)}_{\mathcal{N}=2^{\star}}(m) \equiv  \frac{1}{\mathcal{Z}_{\mathcal{N}=2^{\star}}(m)}\int da \,\left( W_0(a,\lambda) \right)^n \,\text{exp}\left[\displaystyle -\text{tr}a^2+m^2\,\mathcal{M}_{0}+O(m^4)\right]\, \ ,  
\end{align}    
and, the integrated correlator \eqref{IntegratedCorrelatorN4} for $\mathcal{N}=4$ SYM reads
\begin{align}
& \mathcal{I}_{\mathcal{N}=4}^{(n)} \equiv \partial_{m}^2\log \mathcal{W}^{(n)}_{\mathcal{N}=2^{\star}}(m)\Big|_{m=0} = 2\,\frac{\langle  \left(W_0\right)^n\,\mathcal{M}_{0} \rangle_{0} - \langle \left(W_0\right)^n \rangle_{0}\,\langle \mathcal{M}_{0}\rangle_{0}}{\langle \left(W_0\right)^n\rangle_{0}}\, ,
\label{IN4}
\end{align}
where $\langle \cdot \rangle_{0}$ denotes the v.e.v. in the free massless matrix model.

On the other hand, as mentioned in the Introduction, in the case of the $\mathbb{Z}_2$ quiver gauge theory, we can consider a half-BPS Wilson loop in the fundamental representation of the $I$-th node of the quiver. However, it proves more convenient to study untwisted $(+)$ and twisted $(-)$ linear combinations of these operators, which are respectively even or odd under the $\mathbb{Z}_2$ symmetry exchanging the nodes of the quiver. As shown in \cite{Pestun:2007rz,Pini:2023lyo}  they admit the following matrix model representation 
\begin{align}
    W_{\pm}(a,b,\lambda)= \frac{1}{\sqrt{2}}(W(a,\lambda)\pm W(b,\lambda))= \frac{1}{N}\sum_{k=0}^{\infty}\frac{1}{k!}\left(\frac{\lambda}{2N} \right)^{\frac{k}{2}} \;A_k^{\pm}\, \ ,
\label{WPlusAndMinus}    
\end{align}
where $A_{k}^{\pm}$ are $\mathbb{Z}_2$ even/odd matrix model operators, given by
\begin{align}
    A_k^{\pm}= \frac{1}{\sqrt{2}}(\text{tr}a^k \pm \text{tr}b^k) \;.
    \label{A}
\end{align}
It is straightforward to see that the vacuum expectation value of a correlator involving the insertion of an odd number of coincident twisted loops vanishes. Therefore, in the following, we consider only correlators involving $2n_{2}$  twisted Wilson loops and $n_{1}$ untwisted ones. The small mass expansion of this quantity in the mass deformed quiver theory is given by
\begin{align}
& \mathcal{W}^{(n_1,n_2)}_{\mathbb{Z}_2}(m_1,m_2) \equiv \nonumber \\
& \frac{1}{\mathcal{Z}(m_1,m_2)}\int da \,db\, \ \left(W_{+}\right)^{n_1}\; \left(W_{-}\right)^{2n_2}\,\text{exp}\left[\displaystyle  -(\text{tr}a^2+ \text{tr}b^2)-S_0+(m_1^2+m_2^2)\,\mathcal{M}_{\mathbb{Z}_2}+O(m^4)\right]\, \ ,  
\end{align}    
so the integrated correlator \eqref{IntegratedCorrelatorN2}  reads 
\begin{align}
& \mathcal{I}^{(n_1,n_2)} \equiv \partial_{m_i}^2\log \mathcal{W}^{(n_1,n_2)}_{\mathbb{Z}_2}(m_1,m_2)\Big|_{m_i=0} = 2\,\frac{\langle W_{+}^{n_1}\; W_{-}^{2n_2}\,\mathcal{M}_{\mathbb{Z}_2} \rangle - \langle W_{+}^{n_1}\; W_{-}^{2n_2} \rangle\,\langle \mathcal{M}_{\mathbb{Z}_2}\rangle}{\langle  W_{+}^{n_1}\; W_{-}^{2n_2}\rangle} \;.
\label{IZ2}
\end{align}
Here, the vacuum expectation value $\langle \cdot \rangle$ is computed with respect to the massless matrix model of the interacting  $\mathcal{N}=2$ quiver $\mathbb{Z}_2$ gauge theory. For a generic function $f(a,b)$ this is given by
\begin{align}
\langle f(a,b) \rangle = \frac{\langle f(a,b)\,\text{e}^{-S_0} \rangle_0}{\langle \text{e}^{-S_0} \rangle_0} \, \ ,
\label{f}
\end{align}
where $S_0$ is defined in \eqref{S0}. Thus, the problem reduces to a computation within the Gaussian matrix model. Consequently, both the vacuum expectation values in \eqref{IN4} and \eqref{IZ2} can be evaluated using the fusion/fission relations \cite{Billo:2017glv}, together with the identities satisfied by the vacuum expectation values of multiple traces \cite{ Beccaria:2020hgy}
\begin{align}
 t_{p_1,...,p_M}  \equiv \langle \text{tr}a^{\,p_1}\,\cdots\,\text{tr}a^{\,p_M} \rangle_0\, \ .
 \label{tfunction}
\end{align}
An examination of \eqref{IN4} and \eqref{IZ2} shows that the problem of computing these two integrated correlators has been recast into the evaluation of the function \eqref{tfunction} in the $\mathcal{N}=4$ SYM theory, and of $n$-point correlators among the $A_k^{\pm}$ operators in the $\mathcal{N}=2$ theory. In the next section, we review how such computations can be efficiently carried out in both gauge theories.
\subsection{Computations in the large-\texorpdfstring{$N$}{} limit}
\label{subsec:PropertiesAtLargeN}
Let us begin by considering the $\mathcal{N}=4$ SYM theory. As shown for instance in \cite{Beccaria:2020hgy, Billo:2022fnb}, computations can be done more efficiently by performing the change of basis
\begin{align}
\text{tr}a^k = \bigg( \frac{N}{2}  \bigg)^{\frac{k}{2}}\sum_{\ell=0}^{[\frac{k-1}{2}]} \sqrt{k-2\ell} \binom{k}{\ell} \mathcal{P}_{k-2\ell}  + \langle\text{tr}a^k   \rangle_0 \;.
\label{PN4}
\end{align}
The great advantage of the $\mathcal{P}$-basis is the fact that, in the Gaussian matrix model and in the large-$N$ limit, these operators are orthonormal, namely\footnote{Henceforth the symbol $ \simeq$ will denote equality at leading order in the large-$N$ expansion.
}
\begin{align}
\langle \mathcal{P}_k\, \mathcal{P}_q \rangle_0 \, \simeq \, \delta_{k,q}   \;.
\label{Porthonormality}
\end{align}
From the definition \eqref{PN4} it follows that 1-point $\langle \mathcal{P}_{k} \rangle_0$ vanishes. Moreover, as discussed for instance in \cite{Billo:2022fnb}, the planar term of a generic $n$-point correlator among the $\mathcal{P}_k$ operators factorizes à la Wick: into a product of 2-point functions when $n$ is even, or into a product involving a 3-point function and 2-point functions when $n$ is odd. The evaluation of the sub-planar contributions requires more attention. However, for the purposes of this work, it is sufficient to consider only the first non-planar contributions of the 2-, 3-, and 4-point correlators, which has been collected in Appendix \ref{App:Correlators}. 

It also useful to decompose the operator $\mathcal{M}_0$ \eqref{M0} in the $\mathcal{P}$-basis. It can be rewritten as the sum of three terms containing zero, one, and two $\mathcal{P}$ operators, respectively, namely
\begin{align}
\mathcal{M}_0= \mathcal{M}^{(0)}_0 + \mathcal{M}^{(1)}_0+\mathcal{M}^{(2)}_0 \;\;,
\label{M0Decomposition}    
\end{align}
with 
\begin{subequations}
\begin{align}
& \mathcal{M}^{(0)}_0= N^2 \textsf{M}_{0,0}+\textsf{M}_{1,1}-\frac{1}{6}\sum_{k=1}^{\infty}\sqrt{2k+1}\textsf{M}_{1,2k+1} +O(N^{-2})\; \ ,
\label{MM00}\\
&  \mathcal{M}^{(1)}_0= 2N\sum_{k=1}^{\infty}\textsf{M}_{0,2k}\mathcal{P}_{2k} +  \frac{1}{N}\sum_{k=1}^{\infty} \mathcal{Q}_{2k}\mathcal{P}_{2k}  + \frac{1}{N^3}\sum_{k=1}^{\infty}\mathcal{S}_{2k}\mathcal{P}_{2k} + O(N^{-5})\, \ , \label{M01}\\
&\mathcal{M}^{(2)}_0=\sum_{m=2}^{\infty}\sum_{n=2}^{\infty} (-1)^{n-mn}\textsf{M}_{m,n} \;\mathcal{P}_{m} \mathcal{P}_{n} \, \ .
\label{M02}
\end{align}
\label{M0components}
\end{subequations}
It is worth noting that the dependence on the 't Hooft coupling, which appeared as a series in $\lambda$ in \eqref{M0}, can be fully resummed in terms of Bessel functions of the first kind  and it is entirely contained in the following matrix elements
\begin{subequations}
\begin{align}
& \textsf{M}_{0,0}= \int_0^{\infty}dt\,\frac{\text{e}^t\;t}{(\text{e}^t-1)^2} \bigg[1 - \frac{16\pi^2}{\lambda\,t^2} J_1 \bigg(\frac{t \sqrt{\lambda}}{2 \pi} \bigg)J_1 \bigg(\frac{t \sqrt{\lambda}}{2 \pi} \bigg) \bigg] \, ,
\label{M00}
\\[0.5em]
& \textsf{M}_{0,n}= (-1)^{\frac{n}{2}+1} \sqrt{n} \int_0^{\infty}dt\,\frac{\text{e}^t\;t}{(\text{e}^t-1)^2} \bigg(\frac{4\pi}{t \sqrt{\lambda}} \bigg) J_1 \bigg(\frac{t \sqrt{\lambda}}{2 \pi} \bigg)J_n \bigg(\frac{t \sqrt{\lambda}}{2 \pi}   \bigg)\, \ ,
\label{M0n}
\\[0.5em]
&\textsf{M}_{n,m}= (-1)^{\frac{n+m+2nm}{2}+1} \sqrt{nm} \int_0^{\infty}dt\,\frac{\text{e}^t\;t}{(\text{e}^t-1)^2}  J_n \bigg(\frac{t \sqrt{\lambda}}{2 \pi} \bigg)J_m \bigg(\frac{t \sqrt{\lambda}}{2 \pi}   \bigg) \;,
\label{Mnm}
\end{align}
\label{M}
\end{subequations}
and the two coefficients appearing on the right-hand side of \eqref{M01}, which are given by
\begin{align}
\mathcal{Q}_{k} \, = \, & \frac{(-1)^{\frac{k}{2}}}{96\,\pi^2}  \sqrt{k}\int_0^{\infty} dt\;\frac{\text{e}^t\;t}{(\text{e}^t-1)^2} \; \left[\lambda  t^2 J_0\left(\frac{t \sqrt{\lambda }}{2 \pi }\right)-28 \pi  \sqrt{\lambda } t J_1\left(\frac{t \sqrt{\lambda }}{2 \pi }\right)\right] \; J_{k}\left(\frac{t\sqrt{\lambda}}{2\pi} \right)\, \ ,
\label{Qk}\\[0.5em]
\mathcal{S}_{k} \, = \, & \frac{(-1)^{\frac{k}{2}+1}}{737280\,\pi^5}\sqrt{k} \int_0^{\infty}dt\; \frac{\text{e}^t\;t}{(\text{e}^t-1)^2}  \;  \left[\sqrt{\lambda } t \left(-5 \lambda ^2 t^4+4128 \pi ^2 \lambda  t^2-3072 \pi ^4\right) J_1\left(\frac{t \sqrt{\lambda }}{2 \pi }\right) \right. \nonumber \\
& \left. + 96 \pi  \lambda  t^2 \left(8 \pi ^2-3 \lambda  t^2\right) J_0\left(\frac{t \sqrt{\lambda }}{2 \pi }\right) \right]  J_{k}\left(\frac{t\sqrt{\lambda}}{2\pi} \right)\, \, .
\label{Sk}
\end{align}
In particular, the matrix elements \eqref{M} were determined in \cite{Billo:2023ncz}, while the coefficient \eqref{Qk} was obtained in \cite{Pini:2024uia}. Finally, by employing the large-$N$ expansion of the function $ t_{2k}$, namely
\begin{align}
t_{2k} \, = \, & \frac{N^{k+1}}{2^k}\frac{(2k)!}{k!(k+1)!} - \frac{N^{k-1}}{2^{k+1}}\frac{(2k)!}{k!(k-1)!}\left(1 - \frac{k-1}{6} \right) \nonumber \\[0.5em]
& + \frac{N^{k-3}}{2^{k+2}} \frac{(2k)!}{k!(k-2)!} \cdot \frac{1}{360} (5k^3 - 87k^2 + 340k - 12) + O(N^{k-5})\, ,
\label{t2kLargeN}
\end{align}
and applying the method developed in Appendix B of \cite{Pini:2024uia}, we obtained the expression for the final coefficient \eqref{Sk} through a lengthy but straightforward computation.
For future convenience we introduce the matrices $\textsf{M}^{\text{odd}}$ and $\textsf{M}^{\text{even}}$ originating from \eqref{Mnm} and defined as
\begin{align}
(\textsf{M}^{\text{odd}})_{k,\ell} = \textsf{M}_{2k+1,2\ell+1}\, , \qquad (\textsf{M}^{\text{even}})_{k,\ell} = \textsf{M}_{2k,2\ell} \, \ .
\label{MEvenOdd}
\end{align}

The analysis of the large-$ N$ limit of the $ \mathcal{N}=2$ interacting quiver gauge theory can be efficiently carried out in a similar manner. As a first step, it is convenient to perform a change of basis, analogous to \eqref{PN4}, by introducing the operators $ \mathcal{P}_{k}^{\pm}$, defined by the following relation
\cite{Billo:2022lrv,Billo:2022fnb}
\begin{align}
A_k^{\pm}= \bigg( \frac{N}{2}  \bigg)^{\frac{k}{2}}\sum_{\ell=0}^{[\frac{k-1}{2}]} \sqrt{k-2\ell} \binom{k}{\ell} \mathcal{P}_{k-2\ell}^{\pm}  + \langle A_k^{\pm}  \rangle \; \; ,
\label{Pkpm}
\end{align}
where the $A^{\pm}_{k}$ were defined in \eqref{A}.

We observe that, if the interaction action \eqref{S0} were turned off, then the operators $\mathcal{P}_{k}^{\pm}$ would satisfy the orthonormality condition \eqref{Porthonormality} in the free matrix model, and could thus be regarded as the natural extension of the operators $\mathcal{P}_k$ (defined in \eqref{PN4}) to the interacting theory. However, their main importance lies in the fact that the interaction action $S_0$ can be rewritten in a particularly simple form involving only the $ \mathcal{P}_{k}^{-}$ operators, namely \cite{Billo:2022lrv,Billo:2022fnb}
\begin{align}
S_0= - \frac{1}{2}\sum_{k,\ell=1}^{\infty} \mathcal{P}_k^{-}\,\textsf{X}_{k,\ell}\,\mathcal{P}_{\ell}^{-} \;,
\label{S0P}
\end{align}
where the $\textsf{X}$-matrix is given by 
\begin{align}
\textsf{X}_{k,\ell} = 2\;(-1)^{\frac{k+\ell+2k\ell}{2}+1}\sqrt{k\;\ell} \int_0^{\infty} \frac{dt}{t} \;\frac{\text{e}^t}{(\text{e}^t-1)^2} \;J_k\left(\frac{t\sqrt{\lambda}}{2\pi} \right) \;J_{\ell}\left(\frac{t\sqrt{\lambda}}{2\pi} \right) \;,
\end{align}
for $k,\ell \ge 2$ and its entries with opposite parity vanish. We observe that in \eqref{S0P}, all dependence on the coupling $\lambda$ has been resummed in terms of Bessel functions of the first kind, whereas in \eqref{S0} it was expressed as a perturbative series. This resummed form, in turn, enables the derivation of the strong coupling expansion using the techniques discussed in Appendix \ref{app:Analytic}. For later convenience we find it useful to introduce the matrices
$\textsf{X}^{\text{odd}}$ and $\textsf{X}^{\text{even}}$ which are defined as
\begin{align}
(\textsf{X}^{\text{odd}})_{k,\ell} = \textsf{X}_{2k+1,2\ell+1}\, \ , \qquad (\textsf{X}^{\text{even}})_{k,\ell} = \textsf{X}_{2k,2\ell} \, \ .
\label{XOddEven}
\end{align}
Many observables in the large-$N$ limit admit exact expressions in terms of the $\textsf{X}$-matrix. For instance, as shown in \cite{Billo:2022fnb}, the partition function and the free energy $ \mathcal{F}$ of the massless $\mathbb{Z}_2$ quiver gauge theory can be written as
\begin{align}
\mathcal{Z}= (\text{det}[\mathbb{1}-\textsf{X}])^{-\frac{1}{2}}\, \ , \qquad  \mathcal{F}= \frac{1}{2}\text{tr}\log (\mathbb{1}-\textsf{X})\, \ .
\end{align}

Let us now analyze the properties of correlation functions among the $ \mathcal{P}_{k}^{\pm}$ operators in the interacting theory, which will be extensively used in Section \ref{sec:IntegratedCorrelatorZ2}. We begin by observing that, from the definition \eqref{Pkpm}, the 1-point functions vanish trivially, namely  $\langle \mathcal{P}_{k}^{\pm} \rangle = 0$. The first terms in the large-$N$ expansion of the 2-point functions were derived in \cite{Billo:2022fnb, Pini:2024uia} and are given by
\begin{align}
\langle \mathcal{P}_{2k}^{+}\, \mathcal{P}_{2\ell}^{+} \rangle = \delta_{k,\ell} + \frac{\textsf{T}^+_{k,\ell}}{N^2} + O(N^{-4}) \, , \qquad  \langle \mathcal{P}_k^{-}\, \mathcal{P}_\ell^{-} \rangle &\simeq \textsf{D}_{k,\ell}\, \ ,
\label{2pt}
\end{align}
where
\begin{align}
\textsf{T}_{k,\ell}^{+} = \sqrt{k\ell} \left[ \frac{(k^2 + \ell^2 - 1)(k^2 + \ell^2 - 14)}{12} - (k^2 + \ell^2 - 1)\lambda \partial_{\lambda} \mathcal{F} - (\lambda \partial_{\lambda})^2 \mathcal{F} \right] \, ,
\label{Tcoefficient}
\end{align}
and
\begin{align}
\textsf{D}_{k,\ell} = \delta_{k,\ell} + \textsf{X}_{k,\ell} + (\textsf{X}^2)_{k,\ell} + (\textsf{X}^3)_{k,\ell} + \cdots \, ,
\label{D}
\end{align}
which, by using \eqref{XOddEven}, can be divided in its odd and even components
\begin{align}
(\textsf{D}^{\text{odd}})_{k,\ell} = \textsf{D}_{2k+1,2\ell+1}\, \qquad  (\textsf{D}^{\text{even}})_{k,\ell} = \textsf{D}_{2k,2\ell}\, \ .
\label{DOddEven}
\end{align}
The planar terms of the non-trivial 3-point functions are given by \cite{Billo:2022fnb, Pini:2024uia}
\begin{align}
\langle \mathcal{P}^+_{k_1}\, \mathcal{P}^+_{k_2}\, \mathcal{P}^+_{k_3}\rangle \simeq \frac{\sqrt{k_1k_2k_3}}{\sqrt{2}N}\, \ ,
\qquad    \langle \mathcal{P}^+_{k_1} \mathcal{P}^-_{k_2} \mathcal{P}^-_{k_2}\rangle \simeq \frac{\sqrt{k_1}\textsf{d}_{k_2}\textsf{d}_{k_3}}{\sqrt{2}N}\, \ ,  \label{3pt}
  \end{align}
with the understanding that $k_1+k_2+k_3$ is even, and where
\begin{align}
    \textsf{d}_k = \sum_{\ell=0}^{\infty} \sqrt{\ell}\;\textsf{D}_{k,\ell} \;.
\end{align}
Remarkably, as in the case of $\mathcal{N}=4$ SYM theory, the analysis carried out in \cite{Billo:2022fnb} shows that the planar contribution to higher-point functions can still be computed using Wick's theorem. Specifically, a correlation functions among $2n$ $\mathcal{P}^{\pm}$ operators can be decomposed into the sum over all the possible Wick's contractions among  2-point functions \eqref{2pt}. On the other hand, a correlator among $(2n+1)$ $\mathcal{P}^\pm$ operators can be written as the sum over all the possible Wick's contractions of a single 3-point function \eqref{3pt} and $(n-1)$ 2-point functions. 

We finally recall that the operator $\mathcal{M}_{\mathbb{Z}_2}$ \eqref{MZ2} admits the following decomposition in the basis of the $\mathcal{P}_{k}^{\pm}$ operators \cite{Pini:2024uia}
\begin{align}
\label{MZ2Pbasis}
\mathcal{M}_{\mathbb{Z}_2}= \mathcal{M}^{(0)}_{\mathbb{Z}_2} + \mathcal{M}^{(1)}_{\mathbb{Z}_2}+\mathcal{M}^{(2)}_{\mathbb{Z}_2} \;\;,
\end{align}
where the three terms on the r.h.s. contain zero, one and two $\mathcal{P}_{k}^{\pm}$, respectively, and are given by
\begin{subequations}
\begin{align}
\mathcal{M}^{(0)}_{\mathbb{Z}_2} \, = \, & \mathcal{M}^{(0)}_{0}\, \ ,  \\[0.5em] \mathcal{M}^{(1)}_{\mathbb{Z}_2} \, = \, & \sqrt{2}N\sum_{k=1}^{\infty}\textsf{M}_{0,2k}\mathcal{P}_{2k}^+ + \frac{1}{N}\bigg[ \frac{\sqrt{2}\lambda}{32 \pi^2} \sum_{k=1}^{\infty} \textsf{Q}_{0,2k} \mathcal{P}_{2k}^
{+}- \frac{\lambda }{192\pi^2} \sum_{k=1}^{\infty} \textsf{Q}_{2,2k} \mathcal{P}_{2k}^{+} \nonumber \\
& -\lambda \partial_{\lambda}\mathcal{F}\sum_{k=2}^{\infty}\sum_{q=2}^{\infty} \sqrt{k}\;\textsf{M}_{2k,2q}\;\mathcal{P}_{2q}^{+}    \bigg] +O(N^{-3})\, ,
\label{MZ21}
\\[0.5em]
\mathcal{M}^{(2)}_{\mathbb{Z}_2} \, = \, & \frac{1}{2}\sum_{k=2}^{\infty}\sum_{q=2}^{\infty} (-1)^{k-kq}\textsf{M}_{k,q} \; \left(\mathcal{P}_{k}^+ \mathcal{P}_{q}^+ -\mathcal{P}_{k}^- \mathcal{P}_{q}^- \right) \;,
\end{align}
\label{MZ2components}
\end{subequations}
where $\mathcal{M}_{0}^{(0)}$ is given in \eqref{MM00}, the matrix elements $\textsf{M}_{0,k}$ and $\textsf{M}_{k,\ell}$ have been introduced in \eqref{M}. The remaining  coefficients, $\textsf{Q}_{0,2k}$ and $\textsf{Q}_{2,2k}$, appearing on the r.h.s. of \eqref{MZ21}, are defined as follows\footnote{Although it will not be needed in the following, we note that the previously introduced coefficient $\mathcal{Q}_k$  \eqref{Qk} can be expressed in terms of $\textsf{Q}_{0,2k}$ and $\textsf{Q}_{2,2k}$ as
\begin{align}
\mathcal{Q}_{k} =  \frac{\lambda}{16 \pi^2}  \textsf{Q}_{0,2k}  - \frac{\lambda }{96\sqrt{2}\pi^2} \textsf{Q}_{2,2k} \, \ .
\end{align}
}
\begin{subequations}
\begin{align}
& \textsf{Q}_{0,n}= (-1)^{\frac{n}{2}+1} \sqrt{n} \int_0^{\infty}dt\,\frac{\text{e}^t\;t^3}{(\text{e}^t-1)^2} \bigg(\frac{4\pi}{t \sqrt{\lambda}}  \bigg) J_1 \bigg(\frac{t \sqrt{\lambda}}{2 \pi} \bigg)J_n \bigg(\frac{t \sqrt{\lambda}}{2 \pi}\bigg)\, \ , \\[0.5em]
& \textsf{Q}_{n,m}= (-1)^{\frac{n+m+2nm}{2}+1} \sqrt{nm} \int_0^{\infty}dt\,\frac{\text{e}^t\;t^3}{(\text{e}^t-1)^2}  J_n \bigg(\frac{t \sqrt{\lambda}}{2 \pi} \bigg)J_m \bigg(\frac{t \sqrt{\lambda}}{2 \pi}   \bigg) \;.
\end{align}
\label{Q}
\end{subequations}

We are now ready to compute the large-$N$ expansion of the integrated correlators given in \eqref{IN4} and \eqref{IZ2}. In the next section, we begin this analysis by focusing on the $\mathcal{N}=4$ SYM theory.

\section{The integrated correlator \texorpdfstring{$\mathcal{I}^{(n)}_{\mathcal{N}=4}$}{}}
\label{sec:N4}
In this section, we focus on the $\mathcal{N}=4$ SYM theory and we consider the integrated correlator \eqref{IN4}. We first derive the exact expressions, valid for arbitrary values of the 't Hooft coupling, for the first three coefficients of its large-$N$ expansion. Then, using the techniques outlined in Appendix \ref{app:Analytic}, we evaluate the corresponding strong coupling limit. We begin by recalling that, in the $\mathcal{P}$-basis \eqref{PN4}, the Wilson loop \eqref{WN4} takes the following form \cite{Billo:2023ncz} 
\begin{align}
W_0= \langle W_0 \rangle_0+\frac{1}{N}\sum_{k=2}^{\infty} \sqrt{k}\;I_k(\sqrt{\lambda})\;\mathcal{P}_k \, ,
\label{WPbasis}
\end{align}
where $I_{k}(\sqrt{\lambda})$ denotes the modified Bessel function of the first kind. We observe that, by using the expansion \eqref{t2kLargeN} together with the series definition of the Bessel functions $I_k(\sqrt{\lambda})$, the large-$N$ expansion of the v.e.v. appearing on the r.h.s. of \eqref{WPbasis} can be readily evaluated. The result reads  \cite{Erickson:2000af} 
\begin{align}
\langle W_0 \rangle_0 = \textsf{W}^{(L)}_0 + \frac{\textsf{W}^{(NL)}_0}{N^2}+ \frac{\textsf{W}^{(NNL)}_0}{N^4} + O(N^{-6})\;,   
\end{align}
where
\begin{subequations}
\begin{align}
& \textsf{W}^{(L)}_0 = \frac{2}{\sqrt{\lambda}}I_1(\sqrt{\lambda})\, , \\[0.5em]
& \textsf{W}^{(NL)}_0 = \frac{1}{48}\left[\lambda\,I_0(\sqrt{\lambda})-14\sqrt{\lambda}I_1(\sqrt{\lambda})\right]\, , \\[0.5em]
&\,\textsf{W}^{(NNL)}_0 = \frac{1}{46080} \left[\sqrt{\lambda } \left(5 \lambda ^2+1032 \lambda +192\right) I_1(\sqrt{\lambda })-48 \lambda  (3 \lambda +2)I_0(\sqrt{\lambda })\right] \ .
\end{align}
\label{W0components}
\end{subequations}
We are now in the position to consider the integrated correlator \eqref{IN4}, beginning with the large-$N$ expansion of its denominator. To this end, using \eqref{WPbasis} along with the properties of the correlators among the $\mathcal{P}_k$ operators discussed in Section \ref{subsec:PropertiesAtLargeN}, after a lengthy but straightforward computation, we obtain
\begin{align}
\langle \left(W_0\right)^n\rangle_0 = \textsf{W}_0^{(n,0)}+ \frac{\textsf{W}_0^{(n,1)}}{N^2}+ \frac{\textsf{W}_0^{(n,2)}}{N^4} + O(N^{-6})\;,\   
\end{align}
where the $\textsf{W}_0^{(n,i)}$ with $i=0,1,2$ can be expressed in terms of the coefficients \eqref{W0components} and are given by
\begin{subequations}
\begin{align}
\textsf{W}_0^{(n,0)} \, = \, & \left( \textsf{W}^{(L)}_0 \right)^n \, \ , \\[0.5em]
\textsf{W}_0^{(n,1)} \, = \, & n \left( \textsf{W}^{(L)}_0 \right)^{n-2}\bigg[   \textsf{W}^{(L)}_0 \;\textsf{W}^{(NL)}_0  + (n-1)\frac{\sqrt{\lambda}}{4}I_1(\sqrt{\lambda})\;I_2(\sqrt{\lambda})  \bigg]\, \ ,  \\[0.5em]
\textsf{W}_0^{(n,2)} \, = \, & \frac{n}{384}  \left(\textsf{W}_0^{(L)}\right)^{n-4} \bigg[ 384 \left(\textsf{W}_0^{(L)}\right)^3 \textsf{W}_0^{(NNL)} + (n-1)\bigg(12 \lambda  (n-3) (n-2) I_1^2(\sqrt{\lambda})
   I_2^2(\sqrt{\lambda }) \nonumber \\
&   +8 \lambda ^{3/2} (n-2)
   \textsf{W}_0^{(L)} \left(I_1^3(\sqrt{\lambda })+3
   I_2^2(\sqrt{\lambda }) I_1(\sqrt{\lambda
   })\right)+96 \sqrt{\lambda } (n-2) \textsf{W}_0^{(L)}\textsf{W}_0^{(NL)} \times \nonumber \\
& I_1(\sqrt{\lambda }) I_2(\sqrt{\lambda
   })
 +3 \lambda \left(\textsf{W}_0^{(L)}\right)^2 \left(\lambda 
   I_0^2(\sqrt{\lambda })-20 \sqrt{\lambda}
   I_1(\sqrt{\lambda }) I_0(\sqrt{\lambda
   })+(\lambda +28) I_1^2(\sqrt{\lambda
   })\right) \nonumber\\
 &+192 \left(\textsf{W}_0^{(L)}\right)^2 \left(\textsf{W}_0^{(NL)}\right)^2\bigg) \bigg]\;.
\end{align}
\label{DenomComponents}
\end{subequations}
We observe that, upon setting $n = 1$, the expressions above reduce, as expected, to the coefficients \eqref{W0components}. Let us now turn to the numerator of the integrated correlator \eqref{IN4}, which requires more careful treatment. After expressing the operator $ \mathcal{M}_0$ as in \eqref{M0Decomposition}, it can be rewritten as the sum of two connected correlators $\mathcal{K}_i$, namely
\begin{align}
\sum_{i=1}^{2} \mathcal{K}_i \equiv \sum_{i=1}^{2} \langle \left(W_0\right)^n\,\mathcal{M}_0^{(i)}  \rangle_0 - \langle (W_0)^n \rangle_0\, \langle \mathcal{M}_0^{(i)} \rangle_0 \;,
\label{NumeratorN4}
\end{align}
whose large-$N$ expansion takes the form
\begin{align}
 \mathcal{K}_i  =  \mathcal{K}_i ^{(0)} + \frac{ \mathcal{K}_i ^{(1)}}{N^2}+ \frac{ \mathcal{K}_i ^{(2)}}{N^4} + O(N^{-6})\, \   \;.
 \label{ComponentsNum}
\end{align}
The coefficients $\mathcal{K}_i^{(j)}$ for $j = 0, 1, 2$ can be determined using the expression \eqref{WPbasis}, along with the large-$N$ expansions of the correlation functions among the $\mathcal{P}_k$ operators collected in Appendix \ref{App:Correlators}. In particular, the planar contribution to \eqref{NumeratorN4} arises only from $\mathcal{K}_1$, since it is straightforward to verify that $\mathcal{K}_2^{(0)} = 0$. By following the procedure outlined above and making use of the expressions \eqref{M0components}, along with some algebraic manipulations, we obtain \footnote{The notation $\langle \mathcal{P}_{k_1}\mathcal{P}_{k_2}\cdots \mathcal{P}_{k_n} \rangle|_{O(N^{-i})}$ denotes the term of order $O(N^{-i})$ in the large $N$-expansion of the $n$-point function among $\mathcal{P}_k$ operators.} 
\begin{align}
\mathcal{K}_1^{(0)}=&2n\left( \textsf{W}_0^{(L)}\right)^{n-1}\sum_{k=1}^{\infty}\sum_{q=2}^{\infty}\sqrt{q}\;I_q(\sqrt{\lambda}) \;\textsf{M}_{0,2k} \langle \mathcal{P}_q\mathcal{P}_{2k} \rangle \bigg|_{O(N^0)}= \nonumber \\
&2n\left( \textsf{W}_0^{(L)}\right)^{n-1}\sum_{k=1}^{\infty}\sqrt{2k} \;I_{2k}(\sqrt{\lambda})\; \textsf{M}_{0,2k}\, \ . 
\label{K10}
\end{align}
The sum over $k$ can be performed analytically by inserting the explicit expression for the coefficients $ \textsf{M}_{0,2k}$ given in \eqref{M0n}, and by making use of the identity

\begin{align}
\sum_{q=1}^{\infty}(-1)^q(2q)I_{2q}(\sqrt{\lambda})J_{2q}(x)  = -\frac{\sqrt{\lambda}\,x}{2(x^2+\lambda)}\left[\sqrt{\lambda}\,I_0(\sqrt{\lambda})\,J_1(x)-x\,I_1(\sqrt{\lambda})\,J_0(x)\right] \;,
\label{Besselidentity1}
\end{align}
which allows us to finally rewrite \eqref{K10} as  
\begin{align}
\mathcal{K}_1^{(0)} & = \left( \textsf{W}_0^{(L)}\right)^{n-1} \times \nonumber \\
& 8\pi^2n \int_0^{\infty}dt\; \frac{\text{e}^t\;t}{(\text{e}^t-1)^2}\; \frac{1}{4\pi^2+ t^2}\;J_1\bigg( \frac{t\sqrt{\lambda}}{2\pi} \bigg) \bigg[ I_0(\sqrt{\lambda})J_1 \bigg(  \frac{t\sqrt{\lambda}}{2\pi}  \bigg) - \frac{t}{2\pi}    I_1(\sqrt{\lambda})J_0 \bigg( \frac{t\sqrt{\lambda}}{2\pi} \bigg) \bigg]\, \ . \label{Besseleven} 
\end{align}
We observe that, for $n = 1$, the expression above agrees with the results found in \cite{Billo:2023ncz,Pufu:2023vwo}. 

Let us now turn to the next-to-planar order of \eqref{NumeratorN4}, where both connected correlators $ \mathcal{K}_1$ and $ \mathcal{K}_2$ contribute. Specifically, by using the expressions  \eqref{M0components} and \eqref{WPbasis}, and performing some algebraic manipulations, we find that the contribution due to $ \mathcal{K}_1$ is given by
\begin{align}
\mathcal{K}_1^{(1)}\,  = & \, 2n\,\left(\textsf{W}_0^{(L)}\right)^{n-1}\,\sum_{k=1}^{\infty}\sum_{q=2}^{\infty}\sqrt{q}I_q(\sqrt{\lambda})\textsf{M}_{0,2k}\langle \mathcal{P}_{q}\mathcal{P}_{2k}\rangle_0 \bigg|_{O(N^{-2})}  \nonumber\\
&+ 2n(n-1)\left(\textsf{W}_0^{(L)}\right)^{n-2} W_0^{(NL)}\sum_{k=1}^{\infty} \sqrt{2k}\;I_{2k}(\sqrt{\lambda}) \;\textsf{M}_{0,2k} \nonumber \\
& + 2\binom{n}{2}\left(\textsf{W}_0^{(L)}\right)^{n-2}\sum_{q_1=2}^{\infty}\sum_{q_2=2}^{\infty}\sum_{k=1}^{\infty}\sqrt{q_1q_2}I_{q_1}(\sqrt{\lambda})I_{q_2}(\sqrt{\lambda})\textsf{M}_{0,2k}\langle \mathcal{P}_{q_1}\mathcal{P}_{q_2}\mathcal{P}_{2k} \rangle \bigg|_{O(N^{-1})}\nonumber\\
& +2\binom{n}{3}\left(\textsf{W}_0^{(L)}\right)^{n-3}\sum_{q_1=2}^{\infty}\sum_{q_2=2}^{\infty}\sum_{q_3=2}^{\infty}\sqrt{q_1q_2q_3}I_{q_1}(\sqrt{\lambda})I_{q_2}(\sqrt{\lambda})I_{q_3}(\sqrt{\lambda})\textsf{M}_{0,2k}\langle \mathcal{P}_{q_1}\mathcal{P}_{q_2}\mathcal{P}_{q_3}\mathcal{P}_{2k} \rangle_0 \bigg|_{O(N^0)}   \nonumber \\
& + n \left(\textsf{W}_0^{(L)}\right)^{n-1} \sum_{k=2}^{\infty}\sqrt{2k}\; I_{2k}(\sqrt{\lambda})\mathcal{Q}_{2k} \, ,
\label{NLO1}
\end{align}
where the coefficient $\mathcal{Q}_{2k}$ was introduced in \eqref{Qk}. We observe that, after inserting the explicit expressions for the 2-, 3-, and 4-point functions collected in Appendix~\ref{App:Correlators}, we are left with the task of resumming series involving Bessel functions.
Some of these series, such as those appearing in the second, fourth and fifth lines of expression \eqref{NLO1}, can be performed analytically using the identity \eqref{Besselidentity1}. To handle the remaining series, it proves useful to introduce the coefficients
\begin{align}
\mathcal{B}_p = \sum_{k=1}^{\infty}\sqrt{2k}\,\textsf{M}_{0,2k}\,k^p \,,
\label{Bp}
\end{align}
with $p=0,1,2,3\cdots$. Remarkably, by exploiting the techniques developed in Appendix A of \cite{DeLillo:2025hal}, these sums can be evaluated analytically for arbitrary values of $p$. For example, for $p = 4$ we obtain
\begin{align}
\mathcal{B}_4 = \frac{\sqrt{\lambda}}{8\pi^2}\int_0^{\infty}dt\;\frac{\text{e}^t\;t^2}{(\text{e}^t-1)^2}J_1\left(\frac{t \sqrt{\lambda }}{2 \pi }\right)  \left[2 \pi  \; J_0\left(\frac{t \sqrt{\lambda }}{2 \pi }\right) -\sqrt{\lambda}  t J_1\left(\frac{t \sqrt{\lambda }}{2 \pi }\right)\right] \, \ .
\end{align}
Using the coefficients \eqref{Bp}, the sums in the first and third lines of \eqref{NLO1} can also be evaluated explicitly. For instance, the sum in the first line is found to be equal to
\begin{align}
\sum_{k=1}^{\infty}\sum_{q=2}^{\infty}\sqrt{q}I_q(\sqrt{\lambda})\textsf{M}_{0,2k}\langle\mathcal{P}_{q}\mathcal{P}_{2k}\rangle_0 \bigg|_{O(N^{-2})} \, & = \,  \frac{1}{96}\sqrt{\lambda} \left[\mathcal{B}_0 \left((\lambda +28) I_1(\sqrt{\lambda})-14 \sqrt{\lambda } I_0(\sqrt{\lambda })\right) \right. \nonumber \\
& \left. +2\,\mathcal{B}_2 \left(\sqrt{\lambda } I_0(\sqrt{\lambda })-15 I_1(\sqrt{\lambda })\right)+2\,\mathcal{B}_4 I_1(\sqrt{\lambda })\right]\, \ .
\end{align}
Due to its considerable length, we have chosen to provide the complete resummed expression for the coefficient \( \mathcal{K}^{(1)}_1 \) exclusively in the ancillary \texttt{Mathematica} file.
The next-to-planar contribution due to connected correlator $\mathcal{K}_{2}$ can be evaluated in a similar manner and, after some simplifications, takes the form
\begin{align}
\mathcal{K}_2^{(1)} \,= \, & n\left(\textsf{W}_0^{(L)}\right)^{n-1} \left[ \frac{\lambda\, \mathcal{C}_{0,0}}{4}  \;  I_1^2(\sqrt{\lambda})+ \mathcal{D}_{0,0} \left(\frac{\sqrt{\lambda}}{2} I_0(\sqrt{\lambda })-I_1(\sqrt{\lambda })\right)^2\right] \nonumber\\
& + 2\binom{n}{2}\left(\textsf{W}_0^{(L)}\right)^{n-2} \bigg[ \sum_{p,q=1}^{\infty} \sqrt{(2p)(2q)}\;I_{2p}(\sqrt{\lambda})\;I_{2q}(\sqrt{\lambda})\; \textsf{M}_{2p,2q} \nonumber \\ &  +\sum_{p,q=1}^{\infty} \sqrt{(2p+1)(2q+1)}\;I_{2p+1}(\sqrt{\lambda})\;I_{2q+1}(\sqrt{\lambda})\; \textsf{M}_{2p+1,2q+1}  \bigg] \;, 
\end{align}
where we have introduced the further coefficients
\begin{align}
& \mathcal{C}_{i,j} \equiv \sum_{p,q=1}^{\infty} \sqrt{2p}\; p^i \; \sqrt{2q}\; \;q^j \textsf{M}_{2p,2q}\, \ , \\
& \mathcal{D}_{i,j} \equiv \sum_{p,q=1}^{\infty} \sqrt{2p+1}\; (2p+1)^i \; \sqrt{2q+1}\; \;(2q+1)^j \textsf{M}_{2p+1,2q+1}\;.
\end{align}
with $i,j = 0,1,2\cdots$. Importantly, as with the $\mathcal{B}_p$ coefficients discussed above, by exploiting the explicit expressions of the matrix elements $\textsf{M}_{p,q}$ given in \eqref{Mnm} along with the properties of the Bessel functions, the sums over $p$ and $q$ can be again performed analytically. On the other hand,  the series appearing in the second and third lines can also be evaluated analytically using the identity \eqref{Besseleven} and the identity
\begin{align}
\sum_{k=1}^{\infty} (-1)^k \; (2k+1) \;I_{2k+1} \;(\sqrt{\lambda} )\; J_{2k+1}\left(x \right) = -\frac{\sqrt{\lambda}x}{2(x^2+\lambda)}\left[\sqrt{\lambda } I_3(\sqrt{\lambda }) J_2\left(x\right)+ xI_2(\sqrt{\lambda }) J_3\left( x\right) \right]\;.
\label{Besselodd}
\end{align}
In this way, we also obtained an exact expression for the coefficient $ \mathcal{K}_2^{(1)}$, which, due to its length, is provided in the ancillary \texttt{Mathematica} file. We finally observe that, by following the same procedure, exact expressions for the next-to-next-to-planar coefficients $ \mathcal{K}_1^{(2)}$ and $\mathcal{K}_2^{(2)}$ can also be derived. While the computation is lengthy, it remains conceptually straightforward. The final expressions are provided in the ancillary \texttt{Mathematica} file.

We are now ready to evaluate  the large-$N$ expansion of the integrated correlator  \eqref{IN4}, namely 
\begin{align}
\mathcal{I}^{(n)}_{\mathcal{N}=4} = \mathcal{I}_{0,\,\mathcal{N}=4}^{(n)} + \frac{\mathcal{I}_{1,\,\mathcal{N}=4}^{(n)}}{N^2} + \frac{\mathcal{I}_{2,\,\mathcal{N}=4}^{(n)}}{N^4} + \cdots \, \ . 
\label{IN4largeN}
\end{align}
In particular, by using \eqref{Besseleven}, we find that the planar term reads
\begin{align}
\mathcal{I}_{0,\,\mathcal{N}=4}^{(n)}= 
& \frac{8\pi^2\sqrt{\lambda}\,n}{I_1(\sqrt{\lambda})}  \int_0^{\infty}dt\; \frac{\text{e}^t\;t}{(\text{e}^t-1)^2}\; \frac{1}{4\pi^2+ t^2}\;J_1\bigg(\frac{t\sqrt{\lambda}}{2\pi} \bigg) \bigg[ I_0(\sqrt{\lambda})J_1 \bigg(  \frac{t\sqrt{\lambda}}{2\pi}  \bigg) - \frac{t}{2\pi}    I_1(\sqrt{\lambda})J_0 \bigg( \frac{t\sqrt{\lambda}}{2\pi} \bigg) \bigg] \, .
\label{IN4PlanarExplicit}
\end{align}
and it is worth noting that it satisfies the relation
\begin{align}
\mathcal{I}_{0,\,\mathcal{N}=4}^{(n)} = n\,\mathcal{I}_{0,\,\mathcal{N}=4}^{(1)}\;.
\label{IN4Planar}
\end{align}
By exploiting \eqref{DenomComponents} and \eqref{ComponentsNum}, we find that the subplanar coefficients ($q >0$) admit the following closed-form expressions
\begin{align}
\mathcal{I}^{(n)}_{q,\,\mathcal{N}=4} = 2\frac{\mathcal{K}^{(q)}_1+\mathcal{K}^{(q)}_2}{ \textsf{W}_0^{(n,0)} } -\sum_{i=1}^{q}\frac{ \textsf{W}_0^{(n,i)}}{\textsf{W}_0^{(n,0)}}\,\mathcal{I}^{(n)}_{q-i,\,\mathcal{N}=4}\, .
\label{Irecurrence}
\end{align}
We stress that the dependence on $\lambda$ in all the expressions above is exact. This allows, on one hand, an easy derivation of their weak coupling expansion. More interestingly, their strong coupling behavior can also be computed. To this end, by applying the analytical methods presented in Appendix \ref{app:Analytic}, and after a lengthy computation, we finally obtain
\begin{subequations}
\begin{align}
\mathcal{I}_{0,\,\mathcal{N}=4}^{(n)} \, \underset{\lambda \rightarrow \infty}{\sim} & \, n \left[\sqrt{\lambda} + \left(\frac{1}{2}-\frac{\pi^2}{3}\right) + \frac{3}{8\sqrt{\lambda}} + \left(\frac{3\zeta_3}{2}+\frac{3}{8}\right)\frac{1}{\lambda} + \left(\frac{3\zeta_3}{2}+\frac{63}{128}\right)\frac{1}{\lambda^{3/2}} \right. \nonumber
\label{I0N4}\\[0.5em]
& \left. \frac{9}{32\,\lambda^2}(10\zeta_5+7\zeta_3+3) + \frac{9}{1024\,\lambda^{5/2}}(640\zeta_5+384\zeta_3+211) + O\left(\frac{1}{\lambda^{7/2}}\right)  \right]\, \ , \\[0.5em]
\mathcal{I}_{1,\,\mathcal{N}=4}^{(n)} \, \underset{\lambda \rightarrow \infty}{\sim} & \, \frac{\lambda^{3/2}n(2n-3)}{32} + \frac{\sqrt{\lambda}\,n(5-6n)}{256} + \frac{n(5-6n)(4\zeta_3+1)}{128} +  \frac{9\left(7(5-6n)+16\zeta_3(11-8n)\right)}{4096\,\sqrt{\lambda}} \nonumber \\
& + \frac{3n(5-6n)}{256\lambda}(10\zeta_5+7\zeta_3+3) + \frac{15\,n}{32768\,\lambda^{3/2}}\left(960\zeta_5(5-4n) +(384\zeta_3+211)(5-6n)\right) \nonumber \\
& + O\left(\frac{1}{\lambda^{2}}\right)\,
, \ \label{I1n} \\[0.5em]
\mathcal{I}_{2,\,\mathcal{N}=4}^{(n)} \underset{\lambda \to \infty}{\sim}& \frac{\lambda ^3 \left(60 n^3-270 n^2+211 n\right)}{11520}+\frac{\lambda ^{5/2} \left(67 n-60 n^2\right)}{10240}+\frac{\lambda ^2 \left(-15 n^2-n\right)}{7680}+ \nonumber \\
&\frac{\lambda ^{3/2} \left(-3840 n^3 \zeta_3+13440 n^2 \zeta_3+660 n^2-9600 n \zeta_3+143 n\right)}{245760} + O(\lambda)\, \ .
\label{I2n}
\end{align}
\label{FinalResultN4}
\end{subequations}
This is our final result for the strong coupling expansion of the integrated correlator \eqref{IN4}. We observe that by setting $n = 1$ in \eqref{FinalResultN4}, we recover the expression (2.25) of \cite{Pufu:2023vwo}, where the case of a single Wilson loop was considered.

\section{ The integrated correlator \texorpdfstring{$\mathcal{I}^{(n_1,n_2)}$}{}}
\label{sec:IntegratedCorrelatorZ2}
In this section, we compute the large $N$ limit of the integrated correlator \eqref{IZ2} in the $\mathcal{N}=2$ interacting quiver gauge theory. In general, this observable admits the following expansion
\begin{align}
\mathcal{I}^{(n_1,n_2)} = \mathcal{I}_{0}^{(n_1,n_2)} + \frac{\mathcal{I}_{1}^{(n_1,n_2)}}{N^2} + \frac{\mathcal{I}_{2}^{(n_1,n_2)}}{N^4} + \cdots \, \ . 
\label{In1n2LargeN}
\end{align}
We first separately consider integrated correlators involving only insertions of twisted Wilson loops and those involving only untwisted ones. Then, by exploiting the Wick factorization property of correlation functions among the $\mathcal{P}_{k}^{\pm}$
operators discussed in section \ref{subsec:PropertiesAtLargeN}, we derive the result for the general case. In all instances, we repeatedly use the fact that, when expressed in the basis \eqref{Pkpm}, the untwisted/twisted Wilson loops take the following form
\begin{align}
W_{+} \, = \, \langle W_{+} \rangle + \frac{1}{N}\sum_{q=2}^{\infty}\sqrt{q}\,I_{q}(\sqrt{\lambda})\,\mathcal{P}_{q}^{+}\, , \quad \quad W_{-} \, =   \frac{1}{N}\sum_{q=2}^{\infty}\sqrt{q}\,I_{q}(\sqrt{\lambda})\,\mathcal{P}_{q}^{-}\,\;.
\label{WPlusandMinusPbasis}
\end{align}

\subsection{The twisted integrated correlator \texorpdfstring{$\mathcal{I}^{(0,n)}$}{}}
\label{subsec:twistedI}
As a warm-up, let us begin by considering the simplest integrated correlator $\mathcal{I}^{(0,1)}$ involving only two twisted Wilson loops, which corresponds to setting $n_{2}=1$ and $n_{1}=0$ in \eqref{IZ2}. It is explicitly given by
\begin{align}
 \mathcal{I}^{(0,1)} = \frac{\mathcal{K}_{-}^{(1)}+\mathcal{K}_{-}^{(2)}}{\langle W_{-}W_{-}\rangle}\, \ , 
 \label{I01}
\end{align}
where $\mathcal{K}_{-}^{(1)}$ and $\mathcal{K}_{-}^{(2)}$ are given by the following connected correlators
\begin{subequations}
\begin{align}
& \mathcal{K}^{(1)}_{-} \equiv \langle W_{-}\,W_{-}\,\mathcal{M}^{(1)}_{\mathbb{Z}_2}  \rangle - \langle W_{-}\,W_{-} \rangle \langle \mathcal{M}_{\mathbb{Z}_2}^{(1)} \rangle \, ,
\label{M1con}\\[0.5em]
&  \mathcal{K}^{(2)}_{-} \equiv  \langle W_{-}\,W_{-}\,\mathcal{M}^{(2)}_{\mathbb{Z}_2}  \rangle - \langle W_{-}\,W_{-} \rangle \langle \mathcal{M}_{\mathbb{Z}_2}^{(2)} \rangle\, \ .
\label{M2con}
\end{align}
\end{subequations}
Then, using the expressions for the correlation functions among the $\mathcal{P}_{k}^{\pm}$ operators collected in Section \ref{sec:MatrixModel}, we obtain the planar contribution to the expressions above. Specifically, for the first connected correlator, we find
\begin{align}
\mathcal{K}_{-}^{(1)} \, \simeq \, &  - \frac{\textsf{M}_{1,1}}{N^2}\left(\mathcal{V}_{\text{even}}^2 + \mathcal{V}_{\text{odd}}^2\right) \, \ ,
\label{Connected1Exact}
\end{align}
where we introduced
\begin{subequations}
\begin{align}
\mathcal{V}_{\text{even}} = & \sum_{q,p=1}^{\infty}\sqrt{(2q)(2p)}I_{2q}(\sqrt{\lambda})\textsf{D}^{\text{even}}_{q,p}\, \ , 
\label{Veven}
\\
\mathcal{V}_{\text{odd}} = & \sum_{q,p=1}^{\infty}\sqrt{(2q+1)(2p+1)}I_{2q+1}(\sqrt{\lambda})\textsf{D}^{\text{odd}}_{q,p}\, \ ,
\label{Vodd}
\end{align}
\label{VevenAndodd}
\end{subequations}
and $\textsf{D}_{q,p}^{\text{even}}$ and $\textsf{D}_{q,p}^{\text{odd}}$ are defined in \eqref{DOddEven}. For the connected correlator \eqref{M2con}, we obtain instead
\begin{align}
\mathcal{K}^{(2)}_{-} \simeq & -\frac{1}{N^2}\left[\sum_{q_1,q_2=1}^{\infty}\sum_{p_1,p_2=1}^{\infty}\sqrt{(2q_1)(2q_2)}I_{2q_1}(\sqrt{\lambda})I_{2q_2}(\sqrt{\lambda})\,\textsf{D}^{\text{even}}_{q_1,p_1}\,\textsf{M}^{\text{even}}_{p_1,p_2}\,\textsf{D}^{\text{even}}_{p_2,q_2} + \ \right. \nonumber \\
& \left. \sum_{q_1,q_2=1}^{\infty}\sum_{p_1,p_2=1}^{\infty}\sqrt{(2q_1+1)(2q_2+1)}I_{2q_1+1}(\sqrt{\lambda})I_{2q_2+1}(\sqrt{\lambda})\textsf{D}^{\text{odd}}_{q_1,p_1}\,\textsf{M}^{\text{odd}}_{p_1,p_2}\,\textsf{D}^{\text{odd}}_{p_2,q_2} \right]\, \ .
\label{Connected2Exact}
\end{align}
Moreover, by using \eqref{WPlusandMinusPbasis} it's easy to see that the denominator of the expression \eqref{I01} takes the following form
\begin{align}
\langle W_{-}W_{-} \rangle = & \frac{1}{N^2}\sum_{q_1=2}^{\infty}\sum_{q_2=2}^{\infty}\sqrt{q_1\,q_2}\,I_{q_1}(\sqrt{\lambda})\,I_{q_2}(\sqrt{\lambda})\langle \mathcal{P}_{q_1}^{-}\mathcal{P}_{q_2}^{-} \rangle \nonumber \\
& \simeq  \frac{1}{N^2}\sum_{q_1=2}^{\infty}\sum_{q_2=2}^{\infty}\sqrt{q_1\,q_2}\,I_{q_1}(\sqrt{\lambda})\,I_{q_2}(\sqrt{\lambda})\textsf{D}_{q_1,q_2}\, \ ,
\label{Denominator2W}
\end{align}
where we used the large $N$ expansion of the 2-point function \eqref{2pt}.

We remark that  expressions \eqref{Connected1Exact}, \eqref{Connected2Exact} and \eqref{Denominator2W} are exact in the 't Hooft coupling. In the following, we use them to derive the strong coupling expansion of $\mathcal{I}^{(0,1)}$. To this end, we begin by introducing the $\mathcal{N}=4$ SYM connected correlator \cite{Okuyama:2018aij}
\begin{align}
W_{\text{con}}^{(2)} \, \equiv \, \langle W\,W \rangle_0 - \langle W \rangle_0^2 \, \simeq \, \frac{\sqrt{\lambda}}{2\,N^2}I_1(\sqrt{\lambda})I_2(\sqrt{\lambda})\, \ .
\label{W2con}
\end{align}
Next, we rewrite the integrated correlator \eqref{I01} as follows
\begin{align}
\mathcal{I}^{(0,1)} = \frac{W_{\text{con}}^{(2)}}{\langle W_{-}W_{-} \rangle}\times\frac{\mathcal{K}_{-}^{(1)}+\mathcal{K}_{-}^{(2)}}{W_{\text{con}}^{(2)}} \, .    
\end{align}
We then analyze separately the large $\lambda$ expansions of the individual ratios appearing in the expression above. In particular, as shown in \cite{DeSmet:2025mbc}, it holds that
\begin{align}
\frac{\langle W_{-}W_{-} \rangle}{W_{\text{con}}^{(2)}} \, \underset{\lambda \rightarrow \infty}{\sim} \, \frac{\pi^2}{16}\left[1+\frac{2}{\sqrt{\lambda}} + \frac{3-\log(2)}{\lambda}+\left(\frac{15}{4}-3\log(2)-4\log^2(2)\right)\frac{1}{\lambda^{3/2}}+O(\lambda^{-2}) \right]\, \ . 
\label{2ptRatioStrong}
\end{align}
Moreover, the large $\lambda$ expansion of the ratio involving the connected correlator \eqref{Connected1Exact}  can be obtained by exploiting the strong coupling behavior of the expressions \eqref{VevenAndodd}, which is given by \cite{DeSmet:2025mbc}
\begin{subequations}
\begin{align}
& \frac{\mathcal{V}_{\text{odd}}}{I_1(\sqrt{\lambda})}\, \underset{\lambda \rightarrow \infty}{\sim} \, \frac{\pi\,\sqrt{\lambda}}{8} -\frac{3\pi\log(2)}{16\sqrt{\lambda}} - \frac{3\pi\log(2)}{4\,\lambda}\left(\frac{1}{4}+\log(2)\right) + O(\lambda^{-3/2})\, \\
& \frac{\mathcal{V}_{\text{even}}}{I_1(\sqrt{\lambda})}\, \underset{\lambda \rightarrow \infty}{\sim} \, \frac{\pi\,\sqrt{\lambda}}{8} + \frac{\pi}{16}  + \left(\frac{3\pi}{16}+\frac{\pi\log(2)}{16}\right)\frac{1}{\sqrt{\lambda}} + \left(\frac{3\pi}{64}+\frac{3\pi\log(2)}{32}+\frac{\pi\log^2(2)}{4}\right)\frac{1}{\lambda} + O(\lambda^{-3/2})\ \, .
\end{align}
\label{VevenAndOddStrongCoupling}
\end{subequations}
Finally, the strong coupling evaluation of the ratio between the connected correlator \eqref{Connected2Exact} and $W_{\text{con}}^{(2)}$ requires more care. Since we are not aware of any analytical techniques that allow for an exact computation, we have performed the analysis numerically, as detailed in Appendix \ref{app:Numerics}. Here, we simply report the result, which reads
\begin{align}
\frac{\mathcal{K}^{(2)}_{-}}{W_{\text{con}}^{(2)}} \, \underset{\lambda \rightarrow \infty}{\sim} \, \frac{\pi^2(10-\pi^2)}{32}\left[1+\frac{2}{\sqrt{\lambda}} + O(\lambda^{-1})\right]\, \ .
\label{M2exactStrong}
\end{align}
Then, by using the expressions \eqref{VevenAndOddStrongCoupling}, \eqref{M2exactStrong} and the inverse of \eqref{2ptRatioStrong}, we determine the strong coupling limit of the planar term in the expansion \eqref{In1n2LargeN} for the case at hand, namely
\begin{align}
\mathcal{I}^{(0,1)}_0 \, \underset{\lambda \rightarrow \infty}{\sim} \, \sqrt{\lambda} +(8-\pi^2) + \frac{1}{8\sqrt{\lambda}} + O(\lambda^{-1})\, \ .
\label{I0Twisted}
\end{align}
This is our final result for the strong coupling expansion of the planar term of the integrated correlator involving two twisted Wilson loops.

The general case of an integrated correlator involving the insertion of $2n$ twisted Wilson loops can be treated in a completely analogous manner. After a very long but conceptually  straightforward computation, which relies on the Wick factorization property of the correlation functions among the $\mathcal{P}_{k}^{\pm}$ operators, we determine the planar contribution to this observable,  given by
\begin{align} \mathcal{I}^{(0,n)} \simeq \mathcal{I}_0^{(0,n)} = n\,\mathcal{I}_0^{(0,1)} \ .
\label{PlanarOnlyTwisted}
\end{align}
This is our final expression for the integrated correlator involving only twisted Wilson loops.

\subsection{The untwisted integrated correlator \texorpdfstring{$\mathcal{I}^{(n,0)}$}{}}
Let us now move to the case of the integrated correlator $\mathcal{I}^{(n,0)}$, which involves only insertions of untwisted Wilson loops. The case $n=1$ was previously studied in \cite{Pini:2024zwi,DeSmet:2025mbc}. Here, we consider the general case with arbitrary $n$ and compare the results with those for $\mathcal{N}=4$ SYM, found in Section \ref{sec:N4}. We begin by recalling that the large $N$ expansion of v.e.v. of the untwisted Wilson loop in the $\mathbb{Z}_2$ quiver gauge theory reads \cite{Rey:2010ry,Pini:2024zwi}
\begin{align}
\langle W_{+} \rangle = \textsf{W}^{(L)} + \frac{\textsf{W}^{(NL)}}{N^2} + O(N^{-4})\, \   
\end{align}
with 
\begin{align}
& \textsf{W}^{(L)} = \frac{2\sqrt{2}}{\sqrt{\lambda}}I_1(\sqrt{\lambda})\, , \\
& \textsf{W}^{(NL)} = \frac{\sqrt{2}}{48}\left(\lambda\,I_0(\sqrt{\lambda})-14\sqrt{\lambda}I_1(\sqrt{\lambda})\right) - \frac{\lambda^{3/2}\partial_{\lambda}\mathcal{F}}{2\sqrt{2}}I_1(\sqrt{\lambda})\, \ . 
\end{align}
Then, the large $N$ expansion of the denominator of the integrated correlator \eqref{IZ2} is very similar to the corresponding $\mathcal{N}=4$ expansion \eqref{DenomComponents} and it is given by
\begin{align}
\langle \left(W_{+}\right)^n \rangle =\left(\textsf{W}^{(L)}\right)^n  + \frac{n}{N^2}\left(\textsf{W}^{(L)}\right)^{n-2}\left(\textsf{W}^{(L)}\,\textsf{W}^{(NL)} + \frac{(n-1)\sqrt{\lambda}}{4}I_1(\sqrt{\lambda})I_2(\sqrt{\lambda})\right) + O(N^{-4})\, \ .
\label{ExpansionDenominator}
\end{align}
Instead, the numerator reads
\begin{align}
\sum_{i=1}^{2} \langle \left(W_{+}\right)^n\,\mathcal{M}^{(i)}_{\mathbb{Z}_2}  \rangle - \langle (W_{+})^n \rangle \langle \mathcal{M}_{\mathbb{Z}_2}^{(i)} \rangle \, \equiv \sum_{i=1}^{2} \mathcal{K}^{(i)}_{+}\,  .
\label{NumeratorOnlyUntwisted}
\end{align}
Based on the expressions above and by using the properties of the correlation function among $\mathcal{P}_{k}^{\pm}$ operators, after a long computation, we find that
\begin{align}
& \frac{\mathcal{K}_+^{(1)}}{\big(\textsf{W}^{(L)}\big)^n} =   \frac{\sqrt{\lambda}\,n}{2I_1(\sqrt{\lambda})}\sum_{q=1}^{\infty}\sqrt{2q}\,\textsf{M}_{0,2q}\,I_{2q}(\sqrt{\lambda}) \, + \nonumber\\
& \frac{\sqrt{\lambda}\,n}{2\sqrt{2}I_1(\sqrt{\lambda})\,N^2}\left[\sqrt{2}\sum_{k,q=1}^{\infty}\textsf{M}_{0,2k}\sqrt{2q}I_{2q}(\sqrt{\lambda})\textsf{T}^{+}_{q,k} + \frac{\lambda}{32\pi^2}\sum_{q=1}^{\infty}\sqrt{2q}I_{2q}(\sqrt{\lambda})\left(\sqrt{2}\,\textsf{Q}_{0,2q}-\frac{\textsf{Q}_{2,2q}}{6}\right) \nonumber \right. \\
& \left. -\left(\lambda\partial_{\lambda}\mathcal{F}\right)\sum_{k,q=1}^{\infty}\sqrt{2q\,k}\,I_{2q}(\sqrt{\lambda})\,\textsf{M}_{k,q}^{\text{even}} \right] + \frac{\lambda^2n(n-1)(n-2)}{64\,N^2}\frac{I_2(\sqrt{\lambda})}{I_1(\sqrt{\lambda})}\sum_{q=1}^{\infty}\sqrt{2q}I_{2q}(\sqrt{\lambda})\textsf{M}_{0,2q} \nonumber \\
& - \frac{\lambda^2n(n-1)\textsf{M}_{1,1}}{64N^2}\left(1+\frac{I^2_2(\sqrt{\lambda})}{I^2_1(\sqrt{\lambda})}\right) + O(N^{-4})\, \ ,
\label{Ratio1}
\end{align}
We observe that the first line corresponds to the planar contribution, while the subsequent lines encode the next-to-planar ones. Finally, we observe that, at order $O(N^{-2})$, only the terms inside the squared brackets are present also in the $n=1$ case. In a similar way we also obtain
\begin{align}
& \frac{\mathcal{K}^{(2)}_{+}}{\big(\textsf{W}^{(L)}\big)^n} =  \frac{\lambda^2\,n}{256\pi^2\,N^2}\left(\textsf{Q}_{1,1}+\frac{\textsf{Q}_{2,2}}{2}\right) - \frac{n\lambda}{16N^2}\sum_{p,q=1}^{\infty} \left(\textsf{d}_{2p}\,\textsf{M}^{\text{even}}_{p,q}\,\textsf{d}_{2q} + \textsf{d}_{2p+1}\,\textsf{M}^{\text{odd}}_{p,q}\,\textsf{d}_{2q+1}\right) +\nonumber
\\   
& \frac{n(n-1)\lambda}{16\,N^2}\bigg[\sum_{p,q=1}^{\infty}\sqrt{(2p)(2q)}\,\frac{I_{2q}(\sqrt{\lambda})\,I_{2p}(\sqrt{\lambda})}{I_1(\sqrt{\lambda})^2}\textsf{M}^{\text{even}}_{p,q} + \sqrt{(2p+1)(2q+1)}\,\frac{I_{2q+1}(\sqrt{\lambda})\,I_{2p+1}(\sqrt{\lambda})}{I_1(\sqrt{\lambda})^2}\textsf{M}^{\text{odd}}_{p,q}\bigg] \nonumber \\
& + O(N^{-4})\, \ .
\label{Ratio2}
\end{align}
We observe that the second line disappears when the $n=1$ case is considered.  We are now ready to compute the first terms of the large $N$ expansion \eqref{In1n2LargeN} of the integrated correlator $\mathcal{I}^{(n,0)}$.
In particular, the planar contribution is determined only by \eqref{Ratio1}, and reads
\begin{align}
\mathcal{I}_{0}^{(n,0)}  = \frac{\sqrt{\lambda}\,n}{I_1(\sqrt{\lambda})}\sum_{q=1}^{\infty}\sqrt{2q}\,\textsf{M}_{0,2q}\,I_{2q}(\sqrt{\lambda}) = n \,\mathcal{I}^{(1,0)}_0 .
\label{IPlanar}
\end{align}
Interestingly, as found in Section \ref{sec:N4}, the same relation holds for the planar term of the integrated correlator \eqref{IN4} of the $\mathcal{N}=4$ SYM theory. This is consistent with the fact that $\mathcal{I}^{(n,0)}$ is an untwisted observable. On the other hand, the next-to-planar contribution no longer exhibits this property. By applying the analogue of relation \eqref{Irecurrence} to the present case, and performing some algebraic manipulations, we arrive at the following expression
\begin{align}
\mathcal{I}_{1}^{(n,0)} = & 2\;\frac{\mathcal{K}^{(1)}_+ + \mathcal{K}^{(2)}_+}{\big(\textsf{W}^{(L)}\big)^n}\bigg|_{O(N^{-2})} - \left(\frac{\lambda^{3/2}n(n-1)}{32}\frac{I_2(\sqrt{\lambda})}{I_1(\sqrt{\lambda})} - \frac{n\lambda^2}{8}\partial_{\lambda}\mathcal{F}
\right.\nonumber \\
& \left. + \frac{n\sqrt{\lambda}}{96\,I_1(\sqrt{\lambda})}(\lambda\,I_0(\sqrt{\lambda})-14\sqrt{\lambda}I_1(\sqrt{\lambda}))\right)\mathcal{I}_{0}^{(n,0)}\, \ ,
\label{INextToPlanar}
\end{align}
with the understanding that the first term coincide with the contributions of order $O(N^{-2})$ in \eqref{Ratio1} and \eqref{Ratio2}. We observe that both \eqref{IPlanar} and \eqref{INextToPlanar} are exact in the 't Hooft coupling. Their large $\lambda$ expansions can be obtained by following the same steps outlined in \cite{Pini:2024zwi} for $n=1$ and by employing the results collected in Appendix \ref{app:Analytic} and Appendix \ref{app:Numerics}. Following this procedure, after a long computation, one can actually show that
\begin{align}
\mathcal{I}_1^{(n,0)}\, \underset{\lambda \rightarrow \infty}{\sim} & \, -\frac{n\,\lambda^{3/2}}{3072}\left[24n+(n-1)(3\pi^4-62\pi^2+216)\right] - \frac{n\,\lambda}{6144}(n-1)(23\pi^4-134\pi^2-864) \nonumber \\
& + O(\lambda^{1/2})\, \ .
\label{I1strong}
\end{align}
This concludes our derivation of the strong coupling expansion for the next-to-planar contribution of the integrated correlator $\mathcal{I}^{(n,0)}$. It is easy to verify that for $n=1$, the expression \eqref{I1strong} reproduces the strong coupling expansions found in \cite{DeSmet:2025mbc,Pini:2024zwi}.
\subsection{The mixed integrated correlator \texorpdfstring{$\mathcal{I}^{(n_1,n_2)}$}{}}
Finally, let us consider the large $N$ limit of the most general case: an integrated correlator involving $n_1$ untwisted and $2n_2$ twisted Wilson loops. We begin by observing that, upon using the expressions \eqref{IZ2} and \eqref{WPlusandMinusPbasis}, the large $N$ expansion of its denominator simply reads
\begin{align}
\langle \left(W_{-}\right)^{2n_2}\left(W_{+}\right)^{n_1} \rangle \simeq \langle (W_{-})^{2} \rangle^{n_2} \, \langle W_{+} \rangle^{n_1}\, \ .
\label{DenominatorGeneric}
\end{align}
Moreover, the planar contributions of the two connected correlators appearing in the numerator of \eqref{IZ2} can be derived similarly, and read
\begin{align}
    & \sum_{i=1}^{2}\bigg[\langle  W_{-}^{2n_{2}} \;W_{+}^{n_1}\;\mathcal{M}^{(i)}_{{\mathbb{Z}_2}}\rangle-\langle W_{-}^{2n_{2}} \;W_{+}^{n_1} \rangle \langle \mathcal{M}^{(i)}_{{\mathbb{Z}_2}} \rangle \bigg] \simeq \nonumber \\
    & \sum_{i=1}^{2}\bigg[  \langle W_{-}^{2n_2} \rangle \langle W_+^{n_1}\mathcal{M}^{(i)}_{\mathbb{Z}_2}  \rangle+ \langle W_{+}^{n_1} \rangle \langle W_-^{2n_2}\;\mathcal{M}^{(i)}_{\mathbb{Z}_2}  \rangle\bigg] \simeq \nonumber \\ 
    & \sum_{i=1}^{2}\bigg[ n_1 \langle W_{-}^2 \rangle^{n_2} \langle W_+\rangle^{n_1-1} \langle W_+\mathcal{M}^{(i)}_{\mathbb{Z}_2}  \rangle+ n_2\langle W_{+} \rangle^{n_1} \langle W_-^{2}\rangle^{n_2-1}\langle W_{-}W_{-}\mathcal{M}^{(i)}_{\mathbb{Z}_2} \rangle\bigg] \; ,
    \label{Con1andCon2LargeN}
\end{align}
using the Wick factorization property of correlation functions involving the $\mathcal{P}^{\pm}_k$ operators. Then, by combining \eqref{DenominatorGeneric} and \eqref{Con1andCon2LargeN}, and performing some simplifications along with some algebraic manipulations, we finally obtain the planar term of the expansion \eqref{In1n2LargeN}
\begin{align}
\mathcal{I}_0^{(n_1,n_2)} = n_1\, \mathcal{I}_{0}^{(1,0)} + n_2\, \mathcal{I}_0^{(0,1)}\, \ . \label{In1n2Planar}   
\end{align}
Remarkably, this result shows that the planar contribution $\mathcal{I}_{0}^{(n_1,n_2)}$  is given by a  simple linear combination of the planar terms of the  integrated correlator involving a single untwisted Wilson loop and of the integrated correlator involving two twisted Wilson loops, with numerical coefficients $n_1$ and $n_2$ counting the respective number of Wilson loop insertions.

\section{Conclusions}
\label{sec:conclusions}
In this work, we initiated the study of integrated correlators involving the insertion of $n$-coincident Wilson lines and two moment map operators of conformal dimension two, within the $\mathcal{N}=2$ quiver gauge theory arising as a $\mathbb{Z}_2$ orbifold of $\mathcal{N}=4$ SYM, and within the maximally supersymmetric theory itself. By exploiting supersymmetric localization, we analytically derived exact expressions for the leading orders of the large-$N$ expansion of this observable.

Specifically, in the case of $\mathcal{N}=4$ SYM, the planar term of the expansion is given in~\eqref{IN4PlanarExplicit}, whereas the explicit expressions for the first two subplanar coefficients, being rather lengthy, are provided in the ancillary \texttt{Mathematica} file. Then, by applying the analytical techniques reviewed in Appendix~\ref{app:Analytic}, we derive the corresponding strong coupling expansions, which are presented in \eqref{FinalResultN4}. We find that, in the case $n=1$, our results are in full agreement with those of \cite{Pufu:2023vwo}, which investigated the integrated correlator involving the insertion of a single Wilson line. Moreover, these results constitute the starting point for a deeper analysis of this observable. To this end, we plan to investigate, the so-called ``very strong coupling limit" \cite{Binder:2019jwn,Chester:2020dja} of this observable, in which the $\mathcal{N}=4$ complexified coupling constant $\tau$ is kept fixed while $N$ becomes large. In this regime, instanton corrections to the $\mathcal{N}=2^{*}$ mass-deformed partition function are no longer negligible, but they can still be computed following the approach of \cite{Chester:2019jas}. Their explicit evaluation can, in turn, be instrumental in studying the modular properties of this observable, extending the analysis performed in \cite{Dorigoni:2024vrb}, where only the case $n=1$ was considered. Furthermore, we expect that our results could provide a constraint on the Mellin amplitude \cite{Goncalves:2018fwx,Gimenez-Grau:2023fcy} of the corresponding un-integrated defect correlation function and, consequently, be instrumental to determine the first orders of its large-$N$ expansion by following the procedure detailed in~\cite{Pufu:2023vwo}.

On the other hand, in the context of the $\mathcal{N}=2$ quiver gauge theory, our results cover three different cases. The first consists of an integrated correlator with the insertion of only $n$ untwisted Wilson loops. In this case, we obtain exact expressions for the planar and next-to-planar terms of its large-$N$ expansion, given in equations \eqref{IPlanar} and \eqref{INextToPlanar}, respectively. Furthermore, by employing the methods detailed in Appendix \ref{app:Analytic} and Appendix \ref{app:Numerics}, we have also derived the corresponding strong coupling expansions, given by expressions  \eqref{I0N4} and  \eqref{I1strong}, respectively.
We observe that, since this is an untwisted observable, the planar term \eqref{IPlanar} coincides, as expected, with the result found for the $\mathcal{N}=4$ SYM theory in Section \ref{sec:N4}. In contrast, the next-to-planar term differs from that of the maximally supersymmetric theory and extends the results of \cite{DeSmet:2025mbc,Pini:2024zwi}, where only the case $n=1$ was analyzed. The second case, which was considered in Section \ref{subsec:twistedI}, concerns an integrated correlator with the insertion of an even number of twisted Wilson loops only, which does not have a counterpart in $\mathcal{N}=4$ SYM. As a result, already at the planar level, its large-$N$ expansion differs from that of the maximally supersymmetric theory. Finally, the more general case of an integrated correlator involving insertions of both twisted and untwisted Wilson loops has also been considered. Remarkably, after a lengthy analytical computation, we found that the planar term of this observable admits, as given in equation~\eqref{In1n2Planar}, a  very simple expression in terms of the two distinct types of integrated correlators mentioned above.

As noted in the introduction, all SCFTs considered in this work admit a gravity dual, making it natural to investigate the holographic counterpart of the integrated correlators. As a first step in this direction, in \cite{Pufu:2023vwo} it was shown that the $\mathcal{N}=4$ SYM  integrated correlator involving a single Wilson line is dual to the scattering amplitude of a massless string mode from an extended fundamental string (dual to the Wilson line) stretching across $AdS_5$, together with two massless closed string modes originating from insertions of the moment map operators on the $AdS_5$ boundary. This identification crucially relies on the properties of the effective Green-
Schwarz action \cite{GREEN1984367,GREEN1984475,Kallosh:1998ji,Drukker:2000ep} for the fundamental string 
\begin{align}
S_{\text{F}_1} = -T_{\text{F}_1} \int d^2\sigma \sqrt{-\text{det}\, G_{\mu\nu}(X)\partial_{a}X^{\mu}\partial_{b}X^{\nu}} \, + \, \cdots, \, \
\label{SF1action}
\end{align}
where $T_{\text{F}_1}$ is the string tension, $G_{\mu\nu}$ the target space metric, $X^{\mu}$ ($\mu=0,\cdots,9$) the target-space coordinates, and $\sigma^{a}$ ($a=0,1$)  the worldsheet coordinates. Finally, the ellipses encodes the coupling to the NS-NS 2-form $B_{\mu\nu}$ and higher derivative corrections. Expanding the action \eqref{SF1action} in static gauge, $X^{0}=\sigma^{0}$ and  $X^{1}=\sigma^{1}$, around a background metric $G^{(0)}_{\mu\nu}$, with $G_{\mu\nu} = G^{(0)}_{\mu\nu} + h_{\mu\nu}$ and $h_{\mu\nu}$ representing a bulk graviton, yields both the graviton–worldsheet coupling and interaction vertices involving the scalars $X^{i}$ (with $i=2,\cdots,9$) living on the F1 worldsheet. As discussed in \cite{Pufu:2023vwo}, this setup allows one to compute the contributions from both worldsheet and bulk loop diagrams involving propagating gravitons and worldsheet scalars. In turn, these results reproduce the large-$N$ expansion of the integrated correlator in the ’t Hooft limit, thereby establishing the aforementioned interpretation as a scattering process.

In this work, we first consider $n$ coincident Wilson lines in $\mathcal{N}=4$ SYM,  each dual to a fundamental string with the same boundary contour. The planar result \eqref{IN4Planar} is naturally reproduced holographically by considering $n$ copies of the effective action \eqref{SF1action}, reflecting the independence of the fundamental strings at leading order of the large-$N$ expansion.  Beyond the planar limit, however, the expressions \eqref{I1n} and \eqref{I2n} show that the dependence on the number of Wilson lines is encoded in polynomials in $n$, whose degree grows with the order of the large-$N$ expansion. From the holographic perspective, this behavior is expected to arise from modified contributions of both bulk and worldsheet loops relative to the  $n=1$ case, induced by the presence of multiple fundamental strings. The explicit derivation of these contributions lies beyond the scope of the present work.

For the $\mathbb{Z}_2$ quiver gauge theory, one must distinguish between twisted and untwisted Wilson loops. In the case of untwisted Wilson loops, the holographic dual coincides with that of $\mathcal{N}=4$ SYM \cite{Billo:2022fnb}, so the starting point of the holographic analysis remains the effective action \eqref{SF1action}. As a consequence, the planar result \eqref{IPlanar} is simply given by $n$ copies of the $n=1$ expression and coincides with the $\mathcal{N}=4$ SYM result \eqref{In1n2LargeN}. Beyond the planar limit, the difference between the expression \eqref{I1strong} and the corresponding maximally supersymmetric result \eqref{I1n} can be interpreted holographically as arising from modified bulk and worldsheet loop contributions, reflecting the distinct structure of the two SCFTs. In principle, such contributions could be computed by further expanding the effective action \eqref{SF1action}.
The case of twisted Wilson loops is more subtle. As shown by \eqref{PlanarOnlyTwisted}, in the planar limit the contributions of the $n$ fundamental strings can still be treated independently. However, the discrepancy between \eqref{I0Twisted} and the $\mathcal{N}=4$ SYM result \eqref{I0N4} suggests that, in this case, the effective action \eqref{SF1action} does not capture all relevant terms; in particular, it is plausible that the coupling to the NS–NS two-form $B_{\mu\nu}$ should also be included. The explicit derivation of these effects lies beyond the scope of the present work.

In the future, we plan to further investigate the properties of the observable considered in this work within the context of the $\mathcal{N}=2$ quiver gauge theory with an arbitrary number of nodes $M$ \cite{Billo:2021rdb}, which has $M-1$ distinct twisted sectors. It would therefore be interesting to analyze how the integrated correlator with $n$-coincident Wilson lines depends on these sectors. Moreover, we also plan to study the behaviour of this observable in the limit of long quivers \cite{Beccaria:2023qnu,Korchemsky:2025eyc}. Work along these lines is currently in progress.

\vskip 1cm
\noindent {\large {\bf Acknowledgments}}
\vskip 0.2cm
We are very grateful to Marialuisa Frau for carefully reading the manuscript and to Marco Bill\`{o}, Pieter-Jan De Smet, Alberto Lerda and Paolo Vallarino for many relevant discussions. 
AP would like to thank the String Theory group at the University of Turin for their hospitality during the final stages of this work. 

This research is partially supported by the MUR PRIN contract 2020KR4KN2 ``String Theory as a bridge between Gauge Theories and Quantum Gravity'' and by
the INFN project ST\&FI
``String Theory \& Fundamental Interactions''.
The work of AP is supported  by the Deutsche Forschungsgemeinschaft (DFG, German Research Foundation) via the Research Grant ``AdS/CFT beyond the classical supergravity paradigm: Strongly coupled gauge theories and black holes” (project number 511311749). 

\vskip 1cm

\appendix
\section{Strong coupling expansions via analytical methods}
\label{app:Analytic}
In this Appendix, we provide further details on the analytical techniques employed throughout the paper to derive the strong coupling expansions presented in Section \ref{sec:N4} and Section \ref{sec:IntegratedCorrelatorZ2}. To this end, let us consider integrals of the form
\begin{align}
     S_{k,n,m}^{(p)}=  \int_0^{\infty}dt\; \frac{\text{e}^t }{(\text{e}^t-1)^2} \frac{t^p}{(t^2 + 4\pi^2)^k} \; J_n\left( \frac{t\sqrt{\lambda}}{2\pi}  \right) \; J_m\left( \frac{t\sqrt{\lambda}}{2\pi}  \right) \;, \label{doubleJ}
\end{align}
with $k=1,2$. First, we express the product of the two Bessel functions of the first kind in \eqref{doubleJ} using their Mellin-Barnes integral representation, which is given by
\begin{equation}
J_{n}\left(\frac{t\sqrt{\lambda}}{2\pi} \right) J_{m}\left(\frac{t\sqrt{\lambda}}{2\pi} \right)= \int_{-i\infty}^{i \infty} \frac{ds}{2\pi i} \frac{\Gamma(-s)\Gamma(2s+ 1+ m +n)}{\Gamma(s+ m +1) \Gamma(s+n +1) \Gamma(s+m+n +1)} \left( \frac{t\sqrt{\lambda}}{4\pi}   \right)^{2s+m+n} \;.
     \label{MBJJ}
\end{equation}
Inserting \eqref{MBJJ} into \eqref{doubleJ} and switching the order of integration, we find
\begin{align}
S_{k,n,m}^{(p)}= 
&\int_{-i \infty}^{i \infty} \frac{ds}{2\pi i} \frac{\Gamma(-s)\Gamma(2s+ 1+ m +n)}{\Gamma(s+ m +1) \Gamma(s+n +1) \Gamma(s+m+n +1)} \left( \frac{\sqrt{\lambda}}{4\pi}   \right)^{2s+m+n} \nonumber\\
& \times \int_0^{\infty}dt\;   \frac{\text{e}^t }{(\text{e}^t-1)^2} \frac{t^p}{(t^2 + 4\pi^2)^k}  \;t^{2s+n+m} \;.
    \label{Sfinale}
\end{align}
Following \cite{Pufu:2023vwo}, we introduce the functions 
\begin{equation}
    \mathcal{J}_k(s)= \int_0^{\infty}dt  \frac{\text{e}^t \;}{(\text{e}^t-1)^2} \frac{t^{2s+3}}{(4\pi^2 + t^2)^k}\;,
    \label{J}
\end{equation}
which are defined for $\text{Re}(s)>-1$, but can be analytically extended to the entire complex plane repeatedly exploiting  the recurrence relations  
\begin{align}
&\mathcal{J}_1(s) + 4\pi^2 \mathcal{J}_1(s-1)= \int_0^{\infty} dt\frac{\text{e}^t\; }{\left(\text{e}^t-1\right)^2} t^{2 s+1}=  \Gamma(2s+2)\zeta(2s+1) \; ,
\label{Recurrence1}  \\
& \mathcal{J}_2(s) +4\pi^2 \mathcal{J}_2(s-1) = \mathcal{J}_1(s-1) \; .\label{Recurrence2}
\end{align}
Using the definition \eqref{J}, we rewrite \eqref{Sfinale} as 
\begin{align}
    & S_{k,n,m}^{(p)}= \nonumber\\
    &\int_{-i \infty}^{i \infty} \frac{ds}{2\pi i} \frac{\Gamma(-s)\Gamma(2s+ 1+ m +n)}{\Gamma(s+ m +1) \Gamma(s+n +1) \Gamma(s+m+n +1)} \left( \frac{\sqrt{\lambda}}{4\pi}   \right)^{2s+m+n} \mathcal{J}_k\left( s+\frac{n+m+p-3}{2} \right) \;.
    \label{SJ}
\end{align}
In the strong coupling limit, i.e.  $\lambda \gg 1$, the expression \eqref{SJ} receives contributions from poles located on the negative real axis in the complex 
$s$-plane. The function $\mathcal{J}_k$ contributes in two distinct ways: first, through its residues at negative integer values; second, through its negative half-integer values when these correspond to singularities of other factors in the integrand.

Specifically, the values that $\mathcal{J}_k$ assumes at negative half-integer numbers can be deduced iteratively from \eqref{Recurrence1} and \eqref{Recurrence2}, starting from
\begin{equation}
\mathcal{J}_1\left( \frac{1}{2}\right) =  \frac{ \pi ^2}{3} \left(10-\pi ^2\right) \; ,\quad \quad \mathcal{J}_2\left(\frac{1}{2}\right) = \frac{1}{8} \left(\pi^2-4\zeta (3)-5\right)\, \ .
\end{equation}
Additionally, also from \eqref{Recurrence1} and \eqref{Recurrence2}, we derive recurrence relations for the residues of $\mathcal{J}_k$ at $s=-m$, where $m$ is an integer. For instance, in the case $k=1$, the relation reads
\begin{align}
      \text{Res}\; \mathcal{J}_1(s) \bigg|_{s=-m} + 4\pi^2 \; \text{Res} \;\mathcal{J}_1(s) \bigg|_{s=-m-1}  = (-1)^{m+1} \frac{(1-2m) \zeta(2m)}{(2\pi)^{2m}} \;,
\end{align}
with the starting point 
\begin{align}
      \text{Res}\; \mathcal{J}_1(s) \bigg|_{s=0}=0 \;,
\end{align}
which follows from the fact that  $\mathcal{J}_1$ is regular in $s=0$.

As an explicit example, we compute the first orders of the strong coupling expansion of 
\begin{align}
\mathcal{A}=\frac{1}{I_1(\sqrt{\lambda})^2}\sum_{k=1}^{\infty}\sum_{\ell=1}^{\infty}\sqrt{2k}\,I_{2k}(\sqrt{\lambda})\,\textsf{M}_{2k,2\ell}\,\sqrt{2\ell}\,I_{2\ell}(\sqrt{\lambda}) \;,
\label{example}
\end{align}
encountered in the evaluation of the $\mathcal{I}^{(0,1)}$ integrated correlator in section \ref{subsec:twistedI} . Exploiting the identity \eqref{Besselidentity1}
as well as the explicit expression of the matrix elements $\textsf{M}_{2k,2\ell}$ reported in \eqref{Mnm}, the quantity \eqref{example} can be rewritten as the sum of three terms, namely
\begin{align}
    \mathcal{A}= \mathcal{A}_a +\mathcal{A}_b + \mathcal{A}_c \,,
\end{align}
where 
\begin{subequations}
\begin{align}
    & \mathcal{A}_a = - \pi^2 \lambda\; \frac{I_0(\sqrt{\lambda})^2}{I_1(\sqrt{\lambda})^2} \int_0^{\infty} dt\; \frac{\text{e}^t}{(\text{e}^t-1)^2}\;  \frac{ t^3}{(t^2 + 4\pi^2)^2} J_1 \bigg( \frac{t\sqrt{\lambda}}{2\pi} \bigg) J_1 \bigg( \frac{t\sqrt{\lambda}}{2\pi} \bigg)\, \ ,  \\[0.5em]
    &  \mathcal{A}_b=  - \frac{\lambda}{4} \int_0^{\infty} dt\; \frac{\text{e}^t}{(\text{e}^t-1)^2}\;  \frac{ t^5}{(t^2 + 4\pi^2)^2} J_0 \bigg( \frac{t\sqrt{\lambda}}{2\pi} \bigg) J_0 \bigg( \frac{t\sqrt{\lambda}}{2\pi} \bigg)\, , \\[0.5em]
    & \mathcal{A}_c = \pi \;\lambda\; \frac{I_0(\sqrt{\lambda})}{I_1(\sqrt{\lambda})}\; \int_0^{\infty} dt\; \frac{\text{e}^t}{(\text{e}^t-1)^2}\;  \frac{ t^4}{(t^2 + 4\pi^2)^2} J_1 \bigg( \frac{t\sqrt{\lambda}}{2\pi} \bigg) J_0 \bigg( \frac{t\sqrt{\lambda}}{2\pi} \bigg) \;.
\end{align}
\end{subequations}
We observe that all the previous expressions are of the form \eqref{doubleJ} with 
$k=2$. Making use of \eqref{SJ}, we can rewrite them as
\begin{subequations}
\begin{align}
     & \mathcal{A}_a = - \pi^2 \lambda\; \frac{I_0(\sqrt{\lambda})^2}{I_1(\sqrt{\lambda})^2}  \int_{-i \infty}^{i \infty} \frac{ds}{2\pi i} \frac{\Gamma(-s)\Gamma(2s+3)}{\Gamma(s+2)^2\Gamma(s+3)} \left( \frac{\sqrt{\lambda}}{4\pi}   \right)^{2s+2} \mathcal{J}_2\left( s+1 \right)\, \ , \label{Qaf} \\
    &  \mathcal{A}_b=  - \frac{\lambda}{4}  \int_{-i \infty}^{i \infty} \frac{ds}{2\pi i} \frac{\Gamma(-s)\Gamma(2s+ 1)}{\Gamma(s+1)^3} \left( \frac{\sqrt{\lambda}}{4\pi}   \right)^{2s} \mathcal{J}_2\left( s+1 \right)\, \ , \label{Qbf} \\
    & \mathcal{A}_c = \pi \;\lambda\; \frac{I_0(\sqrt{\lambda})}{I_1(\sqrt{\lambda})}\; \int_{-i \infty}^{i \infty} \frac{ds}{2\pi i} \frac{\Gamma(-s)\Gamma(2s+2)}{\Gamma(s +1) \Gamma(s+2)^2} \left( \frac{\sqrt{\lambda}}{4\pi}   \right)^{2s+1} \mathcal{J}_2\left( s+1 \right) \;. \label{Qcf}
\end{align}
\end{subequations}
At strong coupling, \eqref{Qaf}, \eqref{Qbf} and \eqref{Qcf} receive contributions only from the poles of the ratio of the Gamma functions located at the negative half integers. Therefore, we need to evaluate the function $\mathcal{J}_2$ at those points, which can be done by iteratively applying the recurrence relation \eqref{Recurrence2}. Taking all contributions into account, we finally obtain
\begin{align}
    \mathcal{A} \underset{\lambda \to \infty}{\sim} \frac{ \sqrt{\lambda }}{24} \left(9-\pi ^2\right) \nonumber  + \frac{1}{8} \left(\pi ^2-9\right) -\frac{3 \left(\pi ^2-19\right)}{64 \sqrt{\lambda }} + \frac{\pi ^2-3}{32 \lambda } -\frac{105 \left(132 \zeta_3-87+11 \pi^2\right)}{2048 \lambda ^{3/2}} + O(\lambda^{-2}) \;.
\end{align}

\section{Numerical evaluation of the expression \texorpdfstring{\eqref{Connected2Exact}}{}}
\label{app:Numerics}
\begin{figure}[h!]
  \centering
\includegraphics[width=0.98\textwidth]{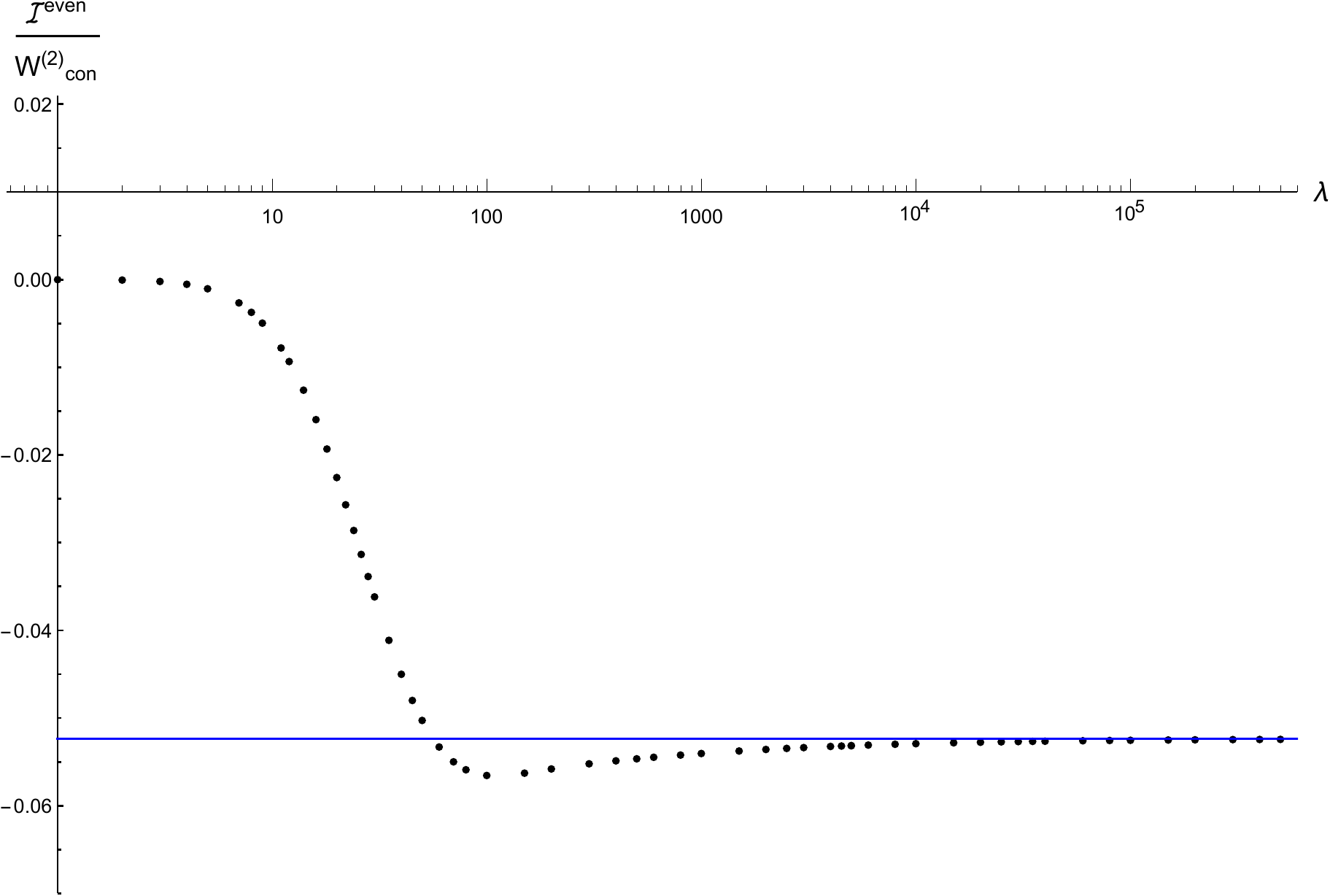}
  \caption{Ratio between $\mathcal{I}^{\text{even}}$ and $W_{\text{con}}^{(2)}$ as a function of the 't Hooft coupling. We observe that this function reaches a plateau for values of $\lambda \geq 10^4$. The blue line represents the value of the coefficient 
 $c_0^{\text{even}}$ obtained by fitting the numerical data (black dots) with the ansatz \eqref{ansatz}. }
  \label{fig:PlotRatio}
\end{figure}

In this appendix, we carry out the numerical evaluation of the leading terms in the strong coupling expansion of expression \eqref{Connected2Exact}, which are required for the computation of the integrated correlator \(\mathcal{I}^{(0,1)}\) discussed in Section~\ref{subsec:twistedI}.
 We use
 the method introduced in \cite{Bobev:2022grf,DeSmet:2025mbc}. As a first step, it is convenient to introduce the quantities
\begin{subequations}
\begin{align}
\widetilde{\textsf{d}}^{\,\,\text{even}}_{p} & = \sum_{q=0}^{\infty}\sqrt{2q}\,I_{2q}(\sqrt{\lambda})\,\textsf{D}^{\text{even}}_{q,p}\, \ ,
\label{dtileEven}\\
\widetilde{\textsf{d}}^{\,\,\text{odd}}_{p} &= \sum_{q=0}^{\infty}\sqrt{(2q+1)}\,I_{2q+1}(\sqrt{\lambda})\,\textsf{D}^{\text{odd}}_{q,p}\, \ ,
\label{dtildeOdd}
\end{align}
\end{subequations}
in terms of which the expression \eqref{Connected2Exact} can be rewritten as
\begin{align}
-\frac{1}{N^2}\sum_{p,q=1}^{\infty}\left(\widetilde{\textsf{d}}^{\, \,\text{even}}_{q}\textsf{M}^{\text{even}}_{q,p}\,\widetilde{\textsf{d}}^{\, \,\text{even}}_{p} + \widetilde{\textsf{d}}^{\, \,\text{odd}}_{q}\textsf{M}^{\text{odd}}_{q,p}\,\widetilde{\textsf{d}}^{\, \,\text{odd}}_{p}\right)\, \ .
\label{TargetNew}
\end{align}
To outline the procedure, and without loss of generality, we find it convenient to focus solely on the even contribution. In this context, we observe that the coefficient \eqref{dtileEven} can be expressed as
\begin{align}
\widetilde{\textsf{d}}^{\,\,\text{even}}_{k}  = \sqrt{2k}\,I_{2k}(\sqrt{\lambda}) - \int_0^{\infty}dt\, U_{k}^{\text{even}}(t)\,Z^{\text{even}}(t) \, ,
\label{dtildeEvenNew}
\end{align}
where $U_k^{\text{even}}(t)$ is defined in equation (2.26) of \cite{DeSmet:2025mbc} and $Z^{\text{even}}(t)$ satisfies the integral equation
\begin{align}
Z^{\text{even}}(t) + 2\int_0^{\infty}ds\,\mathbb{K}^{\text{even}}(t,s)Z^{\text{even}}(s) = \sum_{k=1}^{\infty}\sqrt{2k}I_{2k}(\sqrt{\lambda})U_{k}^{\text{even}}(t)
\label{ZevenEquation}
\end{align}
with the kernel
\begin{align}
\mathbb{K}^{\text{even}}(t,s) = \sum_{k=1}^{\infty}U_k^{\text{even}}(t)U_k^{\text{even}}(s)\, \ .    
\end{align}
By substituting the expression \eqref{dtildeEvenNew} into \eqref{TargetNew} the even contribution becomes
\begin{align} -\sum_{p,q=1}^{\infty}\widetilde{\textsf{d}}^{\, \,\text{even}}_{q}\textsf{M}^{\text{even}}_{q,p}\,\widetilde{\textsf{d}}^{\, \,\text{even}}_{p}  = & -\sum_{k,\ell=1}^{\infty}\sqrt{2k}I_{2k}(\sqrt{\lambda})\textsf{M}_{k,\ell}^{\text{even}}\sqrt{2\ell}\,I_{2\ell}(\sqrt{\lambda}) + \mathcal{I}^{\text{even}}\, \ ,
\label{ExpansionEven}
\end{align}
where
\begin{align}
\mathcal{I}^{\text{even}} \,= & \,  2\sum_{k,\ell=1}^{\infty}\sqrt{2k}\,I_{2k}(\sqrt{\lambda})\,\textsf{M}_{k,\ell}^{\text{even}}\int_0^{\infty}dt\,U_{\ell}^{\text{even}}(t)Z^{\text{even}}(t) \nonumber \\
& - \sum_{k,\ell=1}^{\infty}\textsf{M}_{k,\ell}^{\text{even}}\int_0^{\infty}dt\int_0^{\infty}ds\,U_k^{\text{even}}(t)\,U_{\ell}^{\text{even}}(s)\,Z^{\text{even}}(t)\,Z^{\text{even}}(s)\, \ .
\label{IEven}
\end{align}
Since the strong coupling expansion of the first term on the right-hand side of \eqref{ExpansionEven} was carried out analytically in Appendix~\ref{app:Analytic}, we henceforth focus exclusively on the sum of the remaining terms, denoted\footnote{The analogous quantity for the odd contribution will be denoted by $\mathcal{I}^{\text{odd}}$. } by $\mathcal{I}^{\text{even}}$.
In this regard, it is important to recall that, as shown in \cite{DeSmet:2025mbc}, equations \eqref{dtildeEvenNew}–\eqref{ZevenEquation} can be solved numerically using the Nystr\"om method. This approach enables the subsequent evaluation of the ratio between $\mathcal{I}^{\text{even}}$ and the correlator \eqref{W2con} at large values of the 't Hooft coupling. Specifically, we evaluate this quantity numerically for values of $\lambda$ ranging from $10^4$ to $5\cdot10^5$, with increments of $5\cdot 10^3$. This set of values has been chosen as a compromise between the rapidly increasing computational cost of the algorithm at large values of the coupling and the need to collect data deep in the strong coupling regime. We are confident that this goal is achieved, as the ratio $\mathcal{I}^{\text{even}}/W^{(2)}_{\text{con}}$ clearly approaches a plateau for $\lambda \gg 1$, as illustrated in Figure \ref{fig:PlotRatio}.
We then fit the data using the ansatz
\begin{align}
\sum_{j=0}^{6} c_j^{\text{even}} \lambda^{-j/2} \, ,
\label{ansatz}
\end{align}
where we chose to stop at order $\lambda^{-3}$. This choice enables an accurate determination of the first two numerical coefficients $c_0^{\text{even}}$ and $c_1^{\text{even}}$. Furthermore, we have verified that including higher-order terms in \eqref{ansatz} does not yield any substantial improvement in the quality of the fit.
We perform a similar analysis for the odd contribution in \eqref{TargetNew}. Our findings are reported in Table \ref{tab:Numerical_Coefficients}.
\begin{table}[h!]
\centering
\begin{tabular}{c|c}
 & \text{Estimate} \\
\hline
$c_0^{\text{even}}$ & -0.052358393160(3) \\[0.2em]
$c_1^{\text{even}}$ & -0.058428949(1) \\[0.2em]
$c_0^{\text{odd}}$ & -0.052358392160(0) \\[0.2em]
$c_1^{\text{odd}}$ & 0.138863510(1) \\
\end{tabular}
\caption{Summary of the numerical values obtained for the coefficients of the ansatz \eqref{ansatz} for both the even and the odd contributions. Both systematic and statistical errors have been taken into account.}
\label{tab:Numerical_Coefficients}
\end{table}

Based on these results, we conjecture the form of the first coefficients in the strong coupling expansions
\begin{subequations}
\begin{align}
\frac{\mathcal{I}^{\text{even}}}{W^{(2)}_{\text{con}}} \, \underset{\lambda \rightarrow \infty}{\sim} \, & \frac{1}{192}(144+14\pi^2-3\pi^4) + \frac{1}{128\sqrt{\lambda}}(144+34\pi^2-5\pi^4) + O\left(\frac{1}{\lambda}\right)\, \ ,  \\[0.5em]
\frac{\mathcal{I}^{\text{odd}}}{W^{(2)}_{\text{con}}} \, \underset{\lambda \rightarrow \infty}{\sim} \, & \frac{1}{192}(144+14\pi^2-3\pi^4) - \frac{1}{128\sqrt{\lambda}}(144-46\pi^2+3\pi^4) + O\left(\frac{1}{\lambda}\right)\, \ .
\end{align}
\end{subequations}
As a final step, we sum the two expressions above and, using the method explained in Appendix~\ref{app:Analytic} to perform the summations that can be done analytically (such as the first term on the right-hand side of \eqref{ExpansionEven}), we obtain
\begin{align}
-\frac{1}{W^{(2)}_{\text{con}}}\sum_{p,q=1}^{\infty}\left(\widetilde{\textsf{d}}^{\, \,\text{even}}_{q}\textsf{M}^{\text{even}}_{q,p}\,\widetilde{\textsf{d}}^{\, \,\text{even}}_{p} + \widetilde{\textsf{d}}^{\, \,\text{odd}}_{q}\textsf{M}^{\text{odd}}_{q,p}\,\widetilde{\textsf{d}}^{\, \,\text{odd}}_{p}\right)\, \ \, \underset{\lambda \rightarrow \infty}{\sim}\, \, \frac{\pi^2(10-\pi^2)}{32}\left[1+\frac{2}{\sqrt{\lambda}} + O\left(\frac{1}{\lambda}\right)\right]\, \ .    
\end{align}
This is our final result for the strong coupling expansion of the expression \eqref{Connected2Exact}.

\section{Large-\texorpdfstring{$N$}{} expansions of \texorpdfstring{$\mathcal{N}=4$}{}  correlators}
\label{App:Correlators}
In this Appendix, we collect the large-$N$ expansions of the 2-, 3-, and 4-point correlators among the $\mathcal{P}_k$ operators defined in \eqref{PN4} within $\mathcal{N}=4$ SYM theory, partially extending the results presented in Appendix E of \cite{Billo:2022xas}. These results have been extensively used in Section \ref{sec:N4}. Their derivation strongly relies on the recursion relations, specifically those given by equation (2.30) in \cite{Billo:2017glv}, which allow for the iterative determination of the exact expressions of the function $t_{n_1,\cdots,n_p}$ defined in \eqref{tfunction}. Once these functions are known, the coefficients in the large-$N$ expansions of correlators among the $\mathcal{P}_k$ operators can be obtained by imposing that these quantities are finite-degree polynomials and solving the resulting linear system to determine the coefficients. 

Specifically, for the 2-point function we have to consider separately the case in which the indices of the $\mathcal{P}_k$ operators are both even and both odd. After a long computation we obtained
\begin{align}
\langle \mathcal{P}_{2k}\,\mathcal{P}_{2\ell} \rangle_0 \, = \, &\delta_{k,\ell} +\frac{\sqrt{k\ell}}{12N^2}(k^2 + \ell^2 -1)(k^2 +\ell^2 -14) + \frac{\sqrt{(2k)\,(2\ell)}}{34560\,N^4}\bigg[k^{10}+15 k^8 \ell^2-87 k^8+58 k^6 \ell^4 \nonumber\\
& -1012 k^6 \ell^2+1815 k^6+58 k^4
\ell^6-2074 k^4 \ell^4+13167 k^4 \ell^2-9709 k^4+15 k^2 \ell^8-1012 k^2\ell^6 \nonumber \\
& +13167 k^2 \ell^4 
-23246 k^2 \ell^2+8844 k^2+\ell^{10}-87 \ell^8+1815
   \ell^6-9709 \ell^4+8844 \ell^2-864\bigg] \nonumber \\
& + O(N^{-6})\, \ , \\[0.5em]
\langle \mathcal{P}_{2k+1}\,\mathcal{P}_{2\ell+1} \rangle_0 \, = \, & \delta_{k,\ell} + \frac{\sqrt{(2k+1) (2 \ell+1)}}{24 N^2} \left(k(k+1)+\ell(\ell +1)-14\right)\left(k(k+1)+\ell(\ell +1)\right) \nonumber \\
& +O(N^{-4})\, \ . 
\end{align}
For the 3-point function, to obtain a non-trivial result, the sum of the indices of the $\mathcal{P}_{k}$ operators must be even. Therefore, we consider the following two distinct cases, for which we found
\begin{subequations}
\begin{align}
\langle  \mathcal{P}_{2k_1}\,\mathcal{P}_{2k_2}\,\mathcal{P}_{2k_3} \rangle_0 \, & =\, \frac{\sqrt{2k_1\,2k_2\,2k_3}}{N} + \frac{\sqrt{2k_1\,2k_2\,2k_3}}{72N^3}\bigg[-84+109(k_1^2+k_2^2+k_2^2)\nonumber  
-96(k_1^2k_2^2+k_1^2k_3^3 \nonumber \\
& +k_2^2k_3^2) -26(k_1^4+k_2^4+k_3^4)+12k_1^2k_2^2k_3^2+  +6(k_1^4k_2^2+k_1^4k_3^2+k_2^4k_1^2+k_2^4k_3^2+  k_3^4k_1^2 \nonumber \\
& +k_3^4k_2^2)+ k_1^6+k_2^6+k_3^6\bigg]\, +O(N^{-5})\, .
\end{align}
\begin{align}
\langle \mathcal{P}_{2k_1+2}\mathcal{P}_{2k_2+1}\mathcal{P}_{2k_3} \rangle_0 \, & = \,  \frac{\sqrt{(2k_1+1)\,(2k_2+1)\,2k_3}}{N} + \frac{\sqrt{(2k_1+1)\,(2k_2+1)\,2k_3}}{72N^3}\bigg[84(k_1 + k_2)  + 58(k_1^2 \nonumber \\
& + k_2^2) - 90k_1k_2 + 22k_3^2 -51(k_1^3 + k_2^3) - 84 (k_1^2k_2 + k_2^2k_1) -90 (k_3^2k_1 + k_3^2k_2) \nonumber \\
& - 23 (k_1^4 + k_2^4 + k_3^4) + 
 12(k_1^3k_2 + k_2^3k_1)-78k_1^2k_2^2 - 84 (k_3^2k_2^2 + k_3^2k_1^2) + 12k_1k_2k_3^2 \nonumber \\
 & +3 (k_1^5 + k_2^5) + 6 (k_1^4k_2 + k_2^4k_1 + k_3^4k_1 + k_3^4k_2)+ 12(k_1^3k_2^2 + k_1^2k_2^3 + k_3^2k_2^3 + k_3^2k_1^3) \nonumber \\
 &  + 12 (k_1^2k_2k_3^2 + k_2^2k_1k_3^2) + (k_1^6 + k_2^6 + k_3^6) + 
 12 k_1^2 k_2^2 k_3^2 + 6 (k_1^4 k_2^2 + k_1^2 k_2^4 + k_1^4 k_3^2 \nonumber \\
 & + k_2^4 k_3^2 + k_1^2 k_3^4 + k_2^2 k_3^4)\bigg]\, + O(N^{-5})\, \ .
\end{align}
\label{3point}
\end{subequations}
Finally, we found that the large-$N$ expansion of the 4-point function up to order $O(N^{-4})$ can be written as
\begin{align}
& \langle \mathcal{P}_{k_1}\mathcal{P}_{k_2}\mathcal{P}_{k_3}\mathcal{P}_{k_4} \rangle_0 = \langle \mathcal{P}_{k_1}\mathcal{P}_{k_2}\rangle_0 \langle\mathcal{P}_{k_3}\mathcal{P}_{k_4} \rangle_0 + \langle \mathcal{P}_{k_1}\mathcal{P}_{k_3}\rangle_0 \langle\mathcal{P}_{k_2}\mathcal{P}_{k_4} \rangle_0 + \langle \mathcal{P}_{k_1}\mathcal{P}_{k_4}\rangle_0 \langle\mathcal{P}_{k_2}\mathcal{P}_{k_3} \rangle_0 \nonumber \\
& + \frac{V^{(4)}_{k_1,k_2,k_3,k_4}}{N^2} + O(N^{-4})\, \ ,
\label{4point}
\end{align}
where the vertex $V^{(4)}_{k_1,k_2,k_3,k_4}$ is given by
\begin{align}
V^{(4)}_{k_1,k_2,k_3,k_4} \, = \, & \frac{1}{4}\prod_{i=1}^{4}\sqrt{k_i}\bigg[\sum_{i=1}^{4}(k_i^2-1) -6\big(\delta_{\text{Mod}[k_1\cdot\,k_2,2],0}\delta_{\text{Mod}[k_3\cdot\,k_4,2],1}  \nonumber\\
& +\delta_{\text{Mod}[k_1\cdot\,k_3,2],0}\delta_{\text{Mod}[k_2\cdot\,k_4,2],1}+\delta_{\text{Mod}[k_1\cdot\,k_4,2],0}\delta_{\text{Mod}[k_2\cdot\,k_3,2],1} \nonumber \\
&   +\delta_{\text{Mod}[k_1\cdot\,k_3,2],1}\delta_{\text{Mod}[k_2\cdot\,k_4,2],0}+\delta_{\text{Mod}[k_1\cdot\,k_4,2],1}\delta_{\text{Mod}[k_2\cdot\,k_3,2],0}  \nonumber \\
& + \delta_{\text{Mod}[k_1\cdot\,k_2,2],1}\delta_{\text{Mod}[k_3\cdot\,k_4,2],0} \big)\bigg] \, \ ,
\end{align}
and the Kronecker delta functions ensure that the extra factor $-6$ is present only when exactly two of the indices are odd and the other two are even.

\printbibliography

\end{document}